\documentclass[useAMS,usenatbib]{mn2e}

\usepackage{amssymb}
\usepackage[fleqn]{amsmath}
\usepackage{times}
\usepackage{graphicx}
\usepackage{multirow}
\usepackage{rotating}
\usepackage{longtable,lscape}

\newcommand{\smo}{Smol\v{c}i\'{c}}

\def\f#1   {Fig.~\ref{#1}}
\def\s#1   {Sec.~\ref{#1}}
\def\tab#1   {Tab.~\ref{#1}}
\def\eq#1   {Eq.~\ref{#1}}
\def\t#1   {Tab.~\ref{#1}}
\def\lum   {$\mathrm{L}_\mathrm{1.4GHz}$}
\def\comm#1   {{\tt (COMMENT: #1) }}

\def\msolyr              {$\mathrm{M}_{\odot}\, \mathrm{yr}^{-1}$}

\def\wh                {W~Hz$^{-1}$}

\title[324~MHz VLA-COSMOS survey]
{The VLA-COSMOS Survey.~V.~324~MHz continuum observations}

\author[\smo \ et al. ]{Vernesa Smol{\v c}i{\'c}$^{1}$\thanks{E-mail:vs@phy.hr}, 
Paolo Ciliegi$^{2}$, 
Vibor Jeli\'{c}$^{3,4}$,
Marco Bondi$^{5}$, 
Eva Schinnerer$^{6}$, 
\newauthor Chris L.~Carilli$^{7}$,  
Dominik A.~Riechers$^{8}$, 
Mara Salvato$^{9}$,
Alen Brkovi\'{c}$^{10}$,
\newauthor 
Peter Capak$^{11}$, 
Olivier Ilbert$^{12}$,
Alexander Karim$^{13}$
Henry McCracken$^{14}$,
Nick Z.~Scoville$^{15}$\\
$^{1}$University of Zagreb, Physics Department, Bijeni\v{c}ka cesta 32, 10002 Zagreb, Croatia\\
$^{2}$INAF-Osservatorio Astronomico di Bologna, Via Ranzani 1, I - 40127 Bologna, Italy\\
$^{3}$Kapteyn Astronomical Institute, University of Groningen, PO Box 800, 9700 AV Groningen, the Netherlands\\
$^{4}$ASTRON - the Netherlands Institute for Radio Astronomy, PO Box 2, 7990 AA Dwingeloo, the Netherlands\\
$^{5}$ Istituto di Radioastronomia di Bologna - INAF, via P. Gobetti, 101, 40129, Bologna, Italy \\
$^{6}$Max-Planck-Institut f$\ddot{u}$r Astronomie, K$\ddot{o}$nigstuhl 17, D-69117 Heidelberg, Germany\\
$^{7}$ National Radio Astronomy Observatory, P.O. Box 0, Socorro, NM 87801, USA \\
$^{8}$ Department of Astronomy, Cornell University, Ithaca, New York 14853, USA \\
$^{9}$ Max-Planck-Institut fŸr extraterrestrische Physik, Garching bei MŸnchen, Germany\\
$^{10}$ University of Dubrovnik, Branitelja Dubrovnika 29, 20000 Dubrovnik, Croatia \\
$^{11}$ Spitzer Science Center, 314-6 Caltech, Pasadena, CA 91125, USA\\
$^{12}$ Aix Marseille UniversitŽ, CNRS, LAM (Laboratoire d'Astrophysique de Marseille), UMR 7326, 13388, Marseille, France\\
$^{13}$ Argelander-Institute for Astronomy, Auf dem Hugel 71, 53121 Bonn, Germany\\
$^{14}$ Institut d'Astrophysique de Paris, UMR7095 CNRS, UniversitŽ Pierre et Marie Curie, 98 bis Boulevard Arago, 75014, Paris \\
$^{15}$  California Institute of Technology, MC 249-17, 1200 East California Boulevard, Pasadena, CA 91125, USA
}

\begin{document}


\pagerange{\pageref{firstpage}--\pageref{lastpage}} \pubyear{2011}

\maketitle

\label{firstpage}

\begin{abstract}
We present 90~cm VLA imaging of the COSMOS field, comprising a circular area of 3.14 square degrees at $8.0''\times6.0''$ angular resolution with an average rms of 0.5~mJy/beam. The extracted catalog contains 182 sources (down to 5.5$\sigma$), 30 of which are multi-component sources. Using Monte Carlo artificial source simulations we derive the completeness of the catalog, and we show that our 90~cm source counts agree very well with those from previous studies.
Using X-ray,  NUV-NIR and radio COSMOS data to investigate the population mix of our 90~cm radio sample, we find that our sample is dominated by active galactic nuclei (AGN).
The average 90-20~cm spectral index ($S_\nu\propto\nu^\alpha$, where $S_\nu$ is the flux density at frequency $\nu$, and $\alpha$ the spectral index) of our 90~cm selected sources is -0.70, with an interquartile range of -0.90 to -0.53. 
Only a few ultra-steep-spectrum sources are present in our sample, consistent with results in the literature for similar fields. Our data do not show clear steepening of the spectral index with redshift.
Nevertheless, our sample suggests that sources with spectral indices steeper than -1 all lie at $z\gtrsim1$, in agreement with the idea that ultra-steep-spectrum radio sources may trace intermediate-redshift galaxies (z $\gtrsim$ 1). 
\end{abstract}


\section{Introduction}
\label{sec:intro}

In the last decades optical sky surveys have proven as optimum tools to study properties of galaxies, their formation and evolution. The radio regime is  important  in this context as the observed synchrotron emission from galaxies traces dust-unbiased star formation and active galactic nuclei (AGN). It traces both the high-excitation radio AGN\footnote{High-excitation AGN are defined as those with high-excitation emission lines in their optical spectra, while the spectra of low-excitation AGN are devoid of such lines (Hine \& Longair 1979).} that follow the Unified model for AGN, and low-excitation radio AGN,  not identified as AGN at any other observing wavelength, and inconsistent with the Unified model for AGN \citep[e.g.][]{ho05, hardcastle07, evans06, smo09short}. The radio AGN types exhibit systematic differences in their physical properties; low-excitation radio AGN are hosted by red sequence galaxies with the highest stellar and central supermassive black hole masses, yet the accretion onto their black holes is at sub-Eddington levels \citep{evans06,smo09short,best12}. They are thought to be the population postulated in cosmological models responsible for exerting radio-mode AGN feedback, a heating ingredient in the models necessary to reproduce observed galaxy properties, such as the masses of red galaxies \citep{croton06,bower06,sijacki06}. On the other hand, high-excitation radio AGN are hosted by intermediate mass, 'green valley' galaxies, with intermediate mass supermassive black holes accreting at Eddington levels \citep[e.g.][]{ho05, hardcastle07, evans06, smo09short}.

Although several radio surveys have been obtained at low radio frequencies \citep{bondi07, tasse07, owen09, sirothia09},
to-date the majority of radio surveys have been conducted at 1.4 GHz (20 cm). Opening a new frequency window at lower frequencies offers several advantages, such as detailed radio synchrotron spectral index assessment allowing more precise determination of galaxies' physical properties \citep[e.g.][]{jelic12}, their cosmic evolution \citep[e.g.][]{smo09} and direct identification of intermediate-redshift ($z\gtrsim1$) radio sources via ultra-steep synchrotron spectra \citep[e.g.][]{carlos02}. Compared to frequencies $\geq1.4$~GHz, synchrotron emission at lower frequencies is less affected by synchrotron losses with time (proportional to the square of the frequency), and it has been shown to better correlate with the mechanical power output of radio galaxies into their environment \citep{birzan04, birzan08}, a crucial aspect for studies of radio-mode AGN feedback.  Moreover, compared with other information, the low frequency spectral energy distribution can be used to study the physics of AGN and star-forming galaxies \citep[e.g.][]{birzan04, birzan08, oklopcic10, jelic12}. 

Analyzing their 90 cm data down to a  5$\sigma$ flux limit of 0.35 mJy in the SWIRE $Spitzer$ legacy field, \citet{owen09} found a mean spectral index of $-$0.7, few very steep spectrum 
sources implying the absence of a large population of very steep (and high redshift) radio sources to the limit of their survey and a  flattening of the 90 cm source counts below 3-5 mJy, respectively.   A similar flattening of the 90 cm source counts has been found also by \citet{sirothia09} in the ELAIS N1 field.   This flattening in the radio source counts has been reported earlier at higher frequencies and attributed
to both starburst galaxies and low luminosity AGN \citep[e.g.][]{smo08}. 

Here we present 324 MHz (90~cm) continuum observations of the COSMOS field with the Very Large Array. These low frequency radio data complement the existing 1.4, and ongoing 3 GHz coverage of the COSMOS field at radio wavelengths \citep[][\smo \ et al., in prep.]{schinnerer04, schinnerer07, schinnerer10} providing one of the best fields studied in the radio range and an ideal laboratory for the next generation radio facilities like ASKAP \citep{johnston07}, EMU \citep{norris11} and  the Square Kilometer Array (SKA).  To date the COSMOS  field has been observed with  most major space- and ground-based telescopes over nearly the full electromagnetic spectrum reaching high sensitivities \citep{scoville07, capak07, schinnerer07, taniguchi07, mobasher07, trump07, lilly07, lilly09, hasinger07, elvis09, koekemoer07, sanders07, mccracken12}. 

In \s{sec:obs} \ we describe the observations and data reduction while in \s{sec:catalog} \ and  \s{sec:counts} \  we describe the construction of the catalog and the 90 cm source counts. 
A multiwavelength  analysis is reported in  \s{sec:multi} , while our conclusions are summarized in \s{sec:summary} .

 We define the radio synchrotron spectrum as $S_\nu\propto\nu^\alpha$ where $S_\nu$ is the radio flux density at frequency $\nu$, and $\alpha$ is the spectral index. 

\section{Observations, data reduction and imaging}
\label{sec:obs}

Observations were performed in November 2008 with the VLA in its A configuration. The receivers were tuned to 324~MHz (90~cm, P-band). A single pointing towards the COSMOS field (centered at 10:00:28.6, +02:12:21) was targeted, resulting in a primary beam diameter (FWHM) of 2.3$^\circ$, and a resolution of  $8.0''\times6.0''$  in the final map. A total of 24 hours of observations were scheduled during three nights. Due to the upgrade of some VLA antennas, and the incompatibility of the  P-band receivers with the upgrade, about half of the data was lost, implying a total integration time of $\sim12$~hrs for a 27 antenna array. 

The data were taken in spectral line mode to minimize bandwidth smearing. Two intermediate frequencies (IFs, each with two polarizations) were centered at 321.56 and 326.56~MHz, respectively (we hereafter take the average frequency of 324.1~MHz as the representative one).  A total bandwidth of 3.027~MHz per IF was observed using 31 channels of 97.66~kHz bandwidth each,
resulting in a final bandwidth of 6.027 MHz. 
The source J0521+166 (3C~138), observed at the beginning of each observing run for $\sim13$ minutes, was used for flux and bandpass calibration. The source J1024-008, observed for $1.5$ minutes every $\sim40$~minutes between target observations, was used  for phase and amplitude calibration. The  data reduction was done in AIPS \citep{greisen90}. Bad data were flagged manually (using the AIPS task TVFLG) in each channel before, as well as after calibration to reduce radio frequency interference that could downgrade the quality of the final map. 

After data calibration, the two IFs and two polarizations were imaged separately using a cell size of $1.7''\times1.7''$ (which appropriately samples the CLEAN beam of $8''\times6''$), and 61 separate facets of the field (using the AIPS task SETFC). Twenty more facets were generated using the NVSS catalog to account for bright sources outside the field as these could cause strong side-lobes in the field of interest. The facets were deconvolved via the CLEAN algorithm using the AIPS task IMAGR. CLEAN boxes were set manually around sources. The combined image of the two IFs and two polarizations  was then used as the input model for self-calibration of the data in order to improve the image quality and further reduce the rms in the map.  Self-calibration and imaging as described above were performed multiple times in an iterative manner until satisfactory results were reached. The final imaging was performed on the two IFs and two polarizations  separately using ROBUST=0. The 4 maps were then combined into the final map which was corrected for the primary beam response (using the AIPS task PBCOR with a cut-off at 70\% response limiting the mapped area to $\sim3.14$ square degrees). The final map, shown 
in \f{fig:map} , reaches an angular resolution of $8.0''\times6.0''$ and has an average rms of $\sim0.5$~mJy/beam.

\begin{figure}
\includegraphics[ width=\columnwidth]{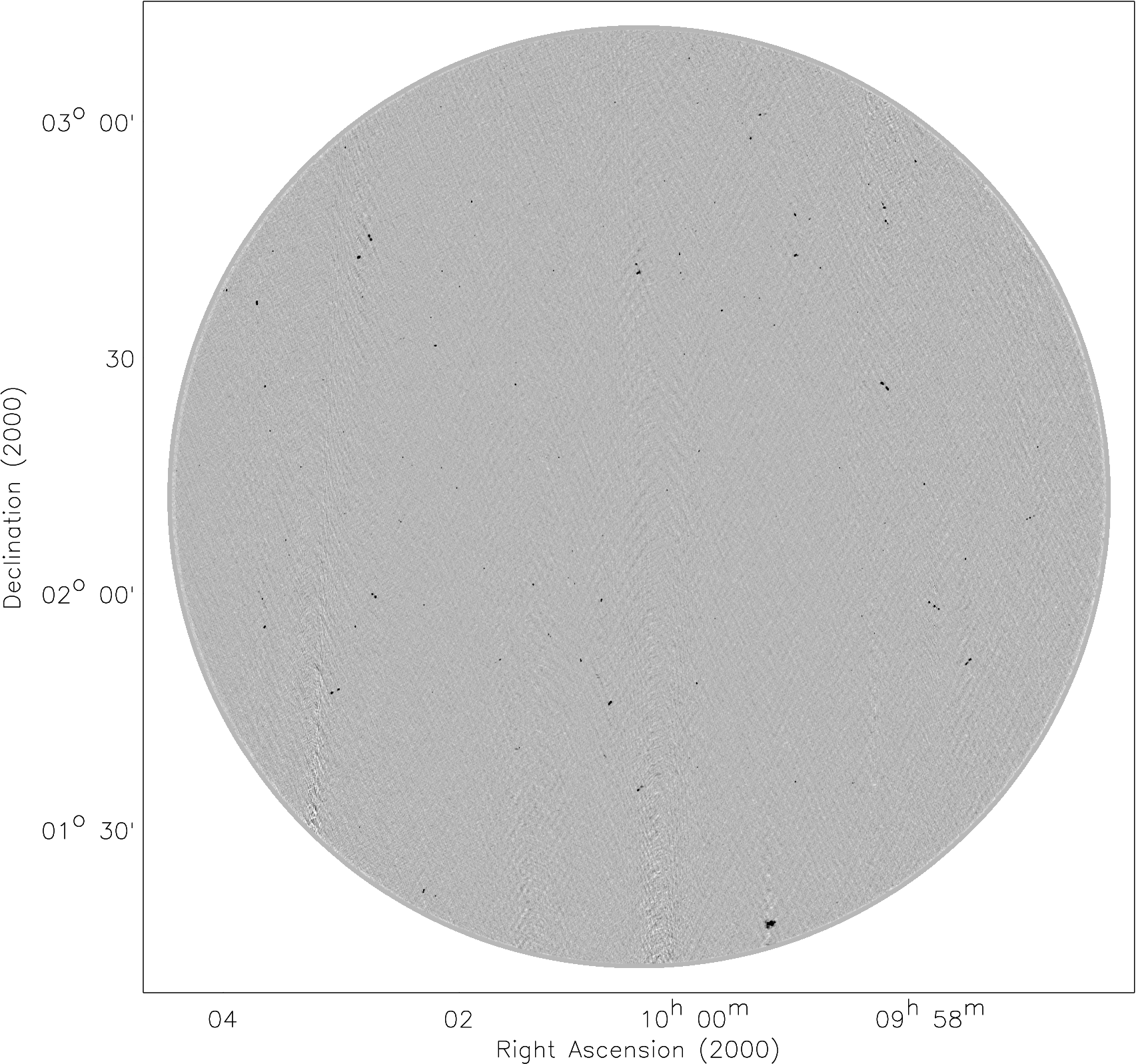}
\caption{ COSMOS field observed at 90 cm.} 
  \label{fig:map}
\end{figure}

\begin{figure}
\includegraphics[ width=\columnwidth]{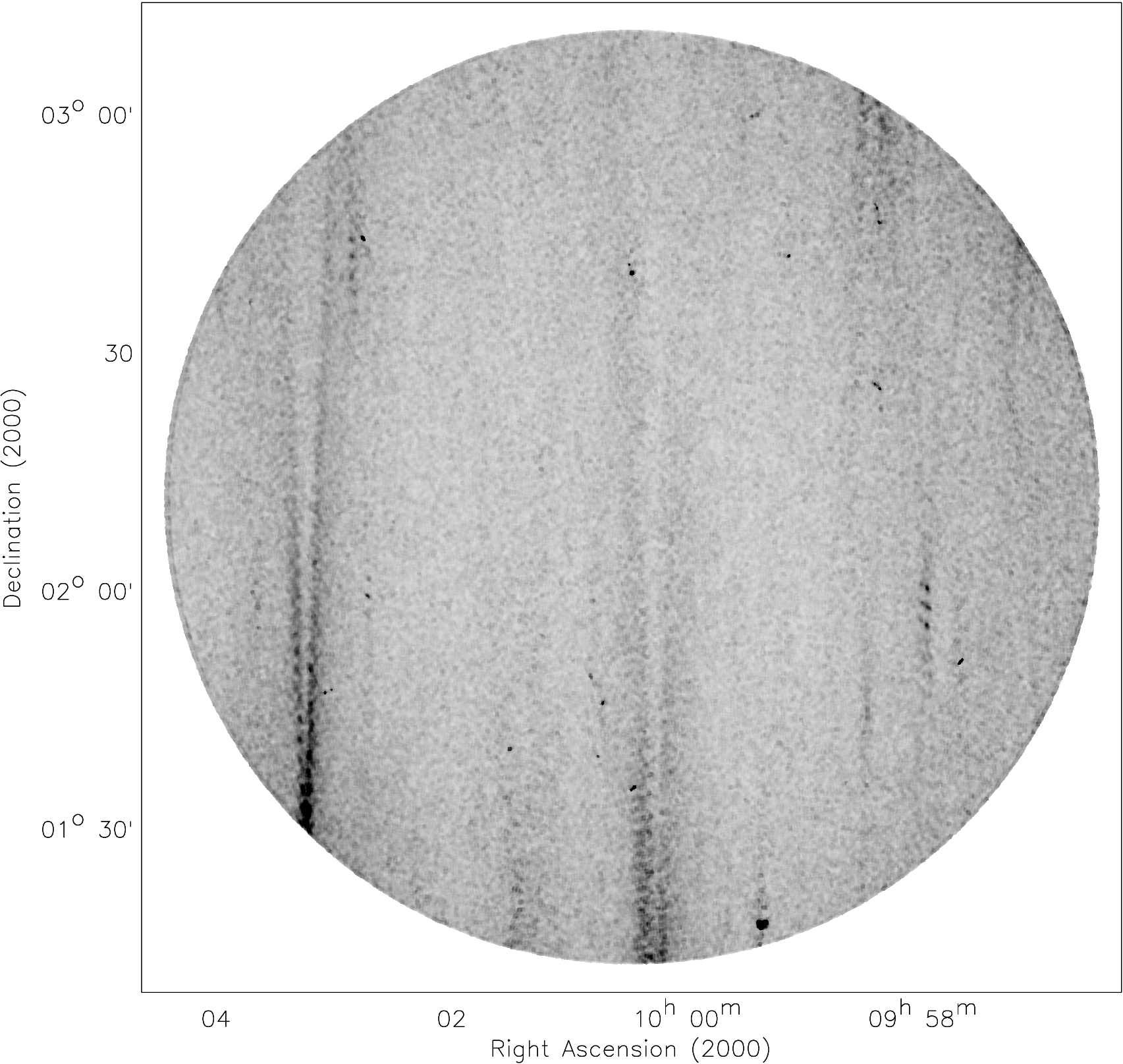}
\caption{ The noise map of the 324 MHz COSMOS field derived via AIPS/RMSD task with a mash size of 20 pixels.}
 \label{fig:noisemap}
\end{figure}

\begin{figure}
\includegraphics[width=\columnwidth]{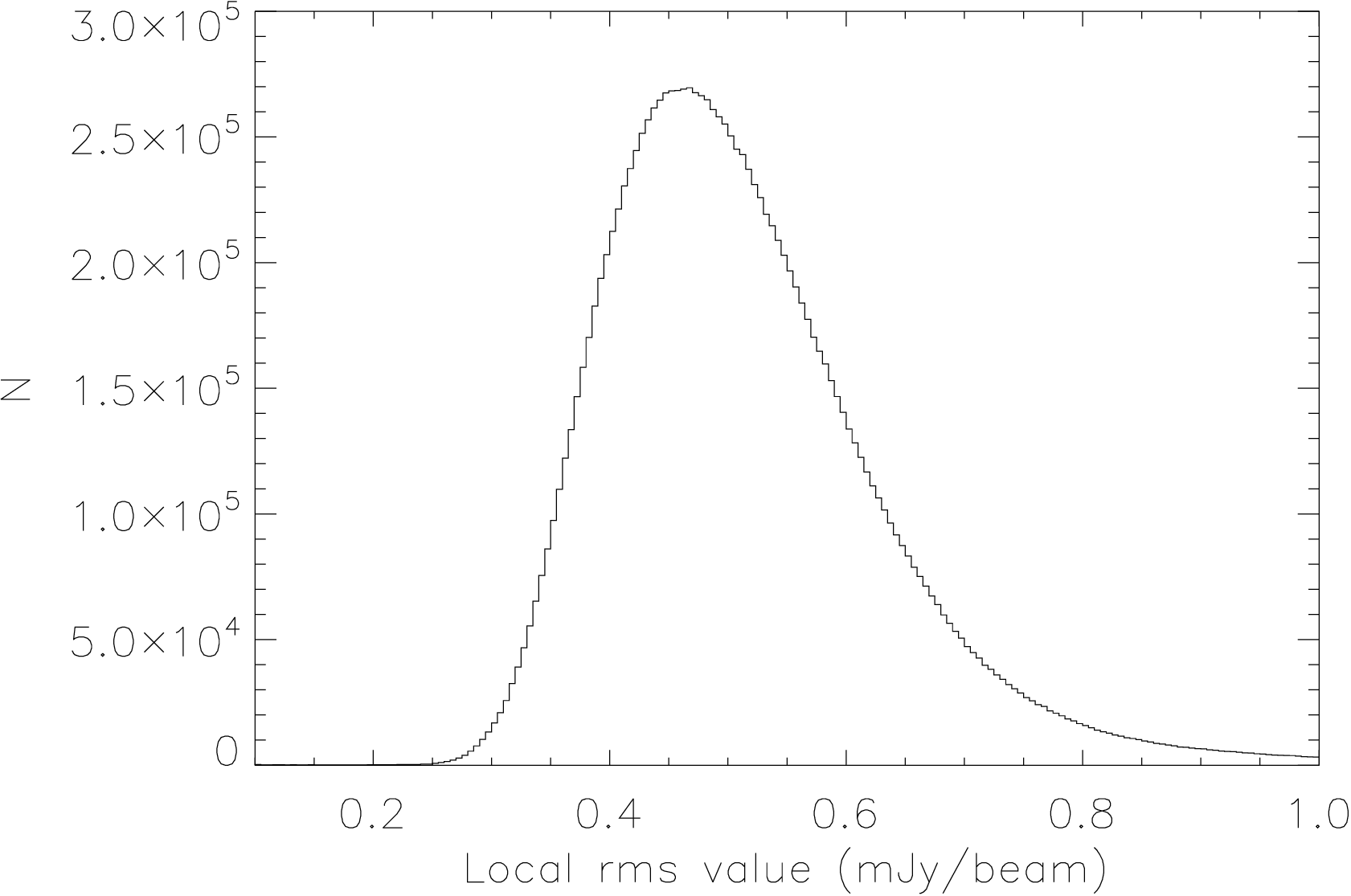}
\caption{ Distribution of the local rms value. The positive tail is due to sources and to  strong side 
 lobes present in the map (see \f{fig:noisemap} ). }
  \label{fig:noisedistr}
\end{figure}

\begin{figure}
\includegraphics[width=\columnwidth]{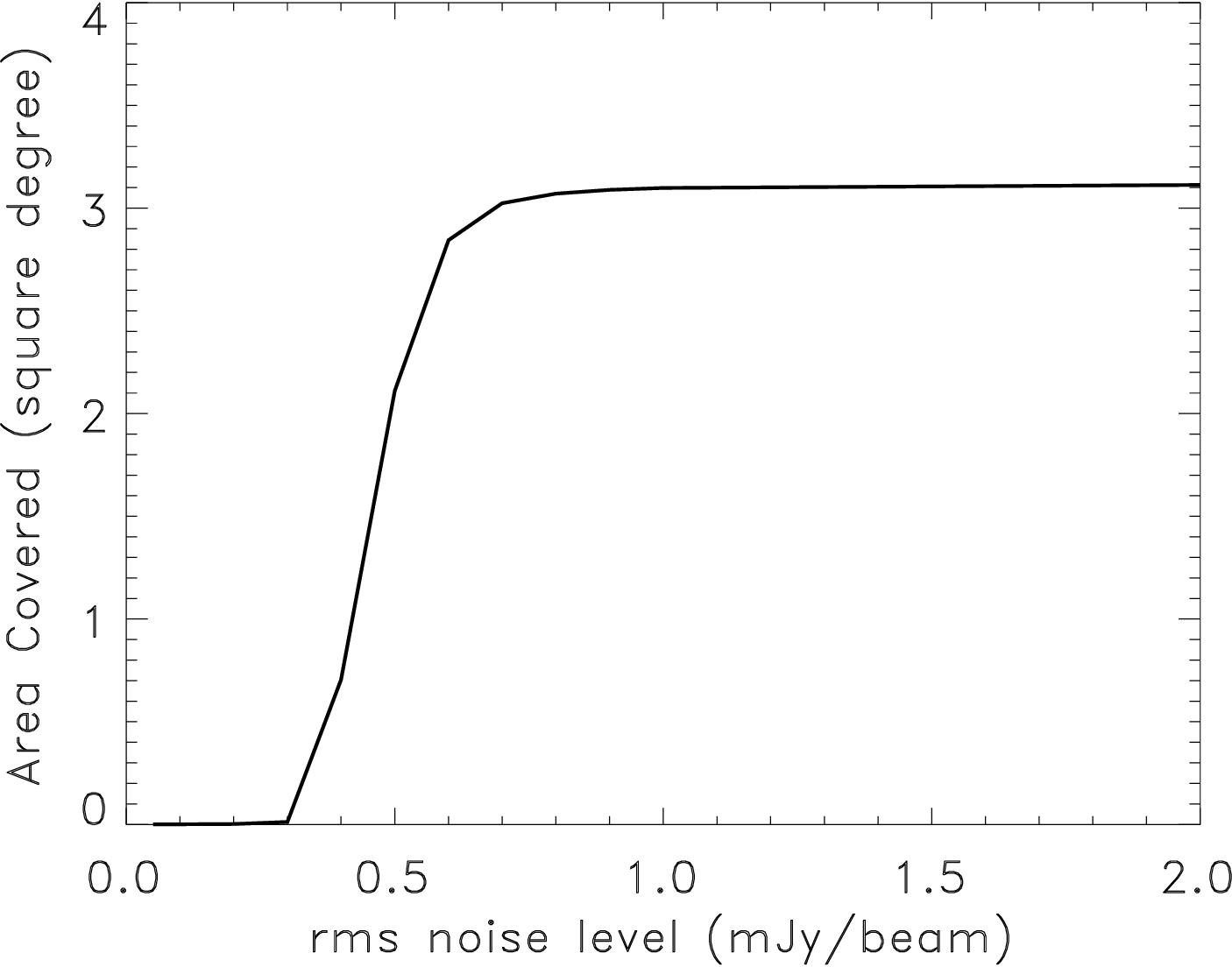}
\caption{ Plot of the rms noise level vs. cumulative area covered. The full area covered is $\sim3.14$ square degrees.} 
  \label{fig:area}
\end{figure}

\section{The 90~cm source catalog: Construction and Completeness}
\label{sec:catalog}

\subsection{Component Extraction: Search and Destroy at 90~cm}
\label{sec:sourceextraction}

The whole image shown in \f{fig:map}  \ (a circular area of radius 1 degree) has been 
used to identify 90\,cm radio components. In order to select a sample 
of sources above a given threshold, defined in terms of local signal-to-noise
ratio, we adopted the  approach already successfully tested during the 
extraction of the VLA-COSMOS 20 cm sources \citep{schinnerer07}. First, the AIPS 
task RMSD was used  to estimate the local background using a mesh size 
of 20 pixels, corresponding  to $34''$. The resulting rms map is shown in \f{fig:noisemap}  \  while 
the distribution of the local rms value is shown in \f{fig:noisedistr}   .  The rms values range
 from about 0.30 mJy/beam 
in the inner regions to about 0.50-0.60 mJy/beam at the edge of the map, with values as high as 
1-2  mJy/beam  in the strong side-lobes present in the map.  The mean rms of the 
full map is  0.51 mJy/beam with  a mode value of 0.46 mJy/beam.  
The cumulative area, as a function of rms is shown in \f{fig:area} . 

Using both the signal and rms maps (\f{fig:map} \ and \f{fig:noisemap} )   as input data we ran the AIPS task SAD to obtain a catalog of candidate components above a given 
local signal-to-noise ratio threshold. The task SAD was run four times with search levels of 10, 8, 6 and 5 in 
signal-to-noise ratio,  using the resulting residual image each time. We recovered all the radio 
components with a local signal-to-noise ratio greater than 5.00.  Subsequently, all the selected components 
have been visually inspected, in order to check their reliability, especially for 
the components near strong side-lobes. After a careful analysis, a signal-to-noise ratio threshold of 5.50
has been adopted as the best compromise between a deep and a reliable catalog. The procedure yielded a total of 246 components with a local signal-to-noise 
ratio greater than 5.50.  

\subsection{Source identification, source sizes and catalog}

More than one component, identified in the 90~cm map as described in the previous section, sometimes belongs to a single radio source  (e.g. large radio galaxies  consist of multiple components). 
Using the 90 cm COSMOS 
radio map we have combined the various components into single sources based on visual inspection. 
The final catalog lists 182 radio sources, 30 of which have been classified as multiple, i.e. they are better 
described by more than a single component.  Moreover, in order to 
ensure a more precise classification,  all  sources identified as  
multi-component sources have been also double-checked using the 20 cm radio map.  We found that
all  the 26 multiple 90 cm radio sources within the 20 cm map have  20 cm sources already classified as multiple.

To generate our final 90~cm source catalog we make use of the VLA-COSMOS Large and Deep Projects over 2 square degrees, reaching down  to an rms of $\sim15~\mu$Jy/beam at 1.4~GHz and $1.5''$ resolution (Schinnerer et al.\ 2007).
The 90 cm COSMOS radio catalog has, however, been extracted from a region of 3.14 square degrees (see \f{fig:map}   ~ and \s{sec:sourceextraction}   ). This implies that a certain  number of 90 cm sources (48)  lie outside the area of the 20 cm COSMOS map used to select the radio catalog.  
Thus, to identify the 20 cm counterparts of the 90 cm radio sources, we used the joint VLA-COSMOS catalog \citep{schinnerer10} for the 134 sources  
within the 20 cm VLA COSMOS area and the VLA - FIRST survey \citep{white97} for the remaining 48 sources.   
The 90 cm sources were cross-matched  with the 20~cm VLA-COSMOS sources using a search radius of 2.5 arcsec, while the cross-match with the VLA-FIRST sources has been done 
using a search radius of 4 arcsec in order to take into account the larger synthesized beam of the VLA - FIRST survey ($\sim$ 5 arcsec).   Finally, all the 90~cm-20~cm associations have beenw
 visually inspected in order to ensure also the association of the  multiple 90 cm radio  sources  for which the value of the search radius used during the cross-match could be too restrictive. 
In summary, out of the total of 182 sources in our 90~cm catalog, 168 have counterparts at 20~cm.

In our 90~cm catalog  multiple component sources are identified by the flag "Mult=1".  
Since the radio core of these multiple sources is not always  detected,
following the same procedure adopted for the 20 cm catalogue,  their {\em virtual}  radio position 
has been calculated as the mean value of the position of all components 
weighted for their total  radio flux.  Moreover, in order to avoid slightly different {\em virtual} 
positions,  for the 26 multiple 90 cm  sources that have been 
visually associated to a 20 cm multiple source (see above) the position has been assumed to be 
coincident with the 20 cm position.  
For all the multiple sources, the total
flux was calculated using the AIPS task TVSTAT, which allows the integration of the map values over 
irregular areas. For these sources the peak flux (at the listed position) is undetermined and
therefore set to a value of -99.999.  Finally, for all the 90 cm single component sources, we report the position 
estimated on the 90 cm map, even for those with a 20 cm counterpart. 

%

In order to determine whether our identified single-component sources are resolved  (i.e. extended, larger than the clean beam)  we make use of the 
ratio between total (S$_{T}$) and peak (S$_{P}$) fluxes (both calculated with the AIPS Gaussian fitting 
algorithm JMFIT  within AIPS/SAD) 
 as this is a direct measure of the extension of a radio source (see \citet{bondi03}).  
  In  \f{fig:res_unres} \ we plot the ratio between the total and the peak  flux
density as a function of the signal to noise ratio S/N (=S$_{P}$/rms) for all the 152 single component sources in the catalog. 
To select the resolved sources, we determined the lower envelope of the points in  \f{fig:res_unres} , which contains 
90\% of the sources with S$_{T}<$S$_P$ and mirrored it above the S$_T$/S$_P$=1 line (upper envelope in \f{fig:res_unres} ).
 We  consider the 56  sources lying above the upper envelope resolved 
which in addition to  the 30 multi component  sources gives a total of 86 resolved sources.  The upper envelope of  \f{fig:res_unres} \ can be characterized  by the equation S$_{T}$/S$_{P}$=1+[100/(S$_{P}$/rms)$^{2.35}$].  
The resolved sources are flagged in the catalog by "Res = 1". 
For the unresolved sources the total flux density is set equal to the peak brightness and the 
angular size is undetermined and set equal to zero in the catalog.

Finally, following the procedure adopted for the VLA COSMOS 20 cm survey, the uncertainties in the peak flux density S$_P$ and integrated flux 
S$_T$ have been calculated using the equation given by   \citet[][see also \citealt{hopkins03, schinnerer04}]{condon97}. For the 
positional uncertainities we used the equations reported in \citet{bondi03} (their eqs. [4] and [5]).

Following  Condon (1997), calibration terms must be  estimated from comparison 
with external data with better accuracy than the one
tested. This is best done using sources strong enough (high S/N), thus yielding the noise terms
in equation 4 and 5  much smaller than the calibration terms.
Our calibration terms have been  calculated from the comparison between 
the position of single component  VLA COSMOS  90 cm sources with S/N$>$10  and the position of  the VLA COSMOS 20 cm counterpart (68  in total). 
The mean values and standard deviations found from this comparison 
are $<$RA$>$=0.22$\pm$0.32 arc sec and $<$DEC$>$=$-$0.27$\pm$0.39 arcsec. These  values 
are consistent with no systematic offset in right ascension and declination and with a calibration term of  
 0.3 arcsec in RA and 0.4 arcsec in DEC.

\begin{figure*}
\vbox to220mm{\vfil
\includegraphics[scale=0.8]{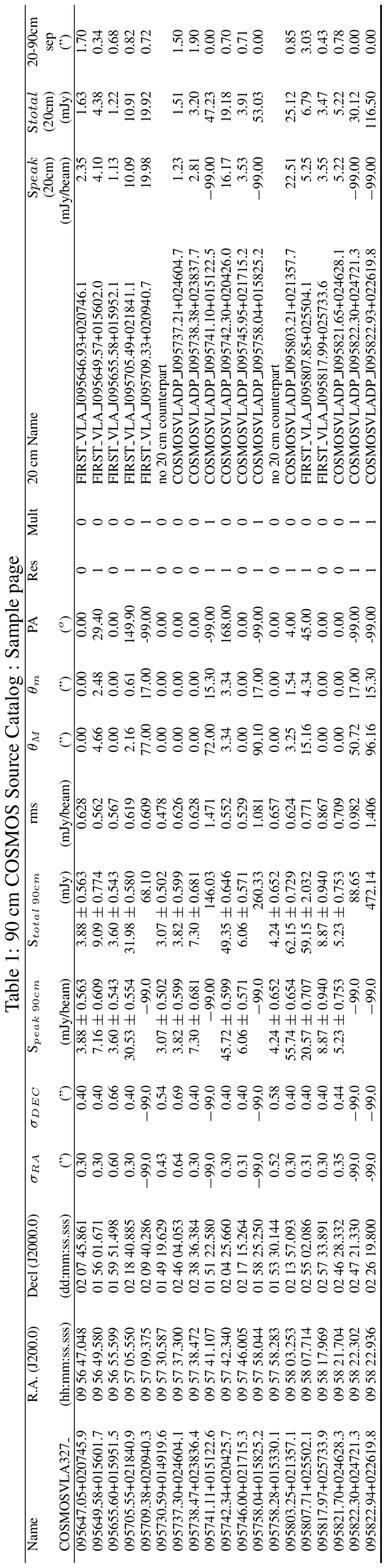}
\vfil}
\end{figure*}

A sample page of the catalog is shown in  Tab.~1.
 For
 each source we report the 90 cm name, the RA and DEC position,  the error on the 
position ($\sigma_{RA}$  and $\sigma_{DEC}$ ), the peak flux with relative error, the total flux with relative error, the 90~cm rms calculated on the 
position of the source,  the deconvolved size of major and minor axes ($\theta_{M}$ and  $\theta_m $) and positional angle (PA) for resolved sources,  the flag 
for resolved (Res = 1) and unresolved (Res = 0) sources, the flag for multiple (Mult = 1) or single (Mult = 0) component sources,  the name of the 20 cm counterpart, 
the 20 cm peak flux, the 20 cm total flux, and the separation between the 90 and 20~cm positions. The full catalog is available in electronic format  at the COSMOS IRSA archive.

\subsection{Comparison with the 20~cm VLA-COSMOS data: 90 cm -- 20 cm spectral indices}
\label{sec:20cm}

The distribution of the 90-20~cm spectral indices is shown in \f{fig:alpha} . We find a median value of $-0.70$ with an interquartile range of -0.90 to -0.53, consistent with the values typically found for radio sources at these flux levels (e.g.\ \citealt{kimball08, owen09}), and further confirming the validity of our 90~cm fluxes.

Fourteen COSMOS 90 cm radio sources do not have a 20 cm counterpart (six of which lie within the 
20 cm  COSMOS area).  The distribution of their 90-20~cm spectral indices,  computed using the VLA-COSMOS and FIRST source detection limits of 75~$\mu$Jy and 1~mJy, respectively, is also shown in \f{fig:alpha} . As expected given the non-detections at 20~cm, these sources have spectral indices steeper than $\alpha\sim-1$. They could be ultra-steep spectrum sources, often located at high redshifts (see Sec.~5.3\  and  \f{fig:specindx} ). However, we cannot exclude the possibility that some of them may also be spurious, specially the seven sources with a relatively low S/N ($<$ 6.5). 





\subsection{Survey Completeness Tests}

In order to estimate the combined effect of noise, source extraction and flux determination on the 
completeness of our sample, following the method described in Bondi et al.\ (2003)  we constructed simulated samples of radio sources down to a flux level of 0.9 mJy, 
i.e. more than a factor of 3 lower than the minimum flux we used to derive the source counts (see next section). 
This allows us to account for those sources with an intrinsic flux below the detection limit which, because of 
positive noise fluctuations, might have a measured flux above the limit. The total flux distribution has 
been simulated using the 90 cm source counts and angular size distribution reported by \citet{owen09}.
In particular, the sizes of sources have been  randomly extracted from three different  normal distributions 
with a mean value of 17, 7 and 4.5 arcsec  for sources with a simulated flux of $>$ 100 mJy,  10-100 mJy and 
$<$10 mJy respectively (see Table 3 in \citealt{owen09}). 

 A total of 10 samples were simulated over an area of 3.14 square degree, i.e.\ equal to the area used to extract 
the 90 cm COSMOS catalog.  According  to the source counts distribution reported by \citet{owen09} (see their Table 8), we expect $\sim$ 195 sources/deg$^2$ with  a flux greater than 0.9 mJy, corresponding to $\sim$ 610
sources for each simulated catalog of 3.14 square degrees.  Considering all the samples, we simulated a total of 6115 
sources, with a flux  distribution as shown in \f{fig:simflux} . 

For each simulated sample, the sources 
were randomly injected in the residual sky image (i.e. in the map with the real sources removed) and were 
recovered with fluxes  measured using the same procedure adopted for the real sources (see Sect. 3.1). 
All the detected simulated sources from all the 10 simulated samples were then binned using the same flux intervals
used in the source counts calculation. The results of 
our simulation are summarized in Table 2, where for each flux density bin we report the number of sources injected in the simulations,  the number of sources detected using the same procedure adopted for the real data and 
the correction factor (C, with the relative standard deviation $\sigma_{C}$) to be applied to our observed source counts.  Our simulations tell us that we are 
strongly incomplete below 3.0 mJy (S/N $\sim$ 6.0)  while  at higher flux levels  we are 
missing sources ($\sim$ 10\%) only in the  bin 3.0-4.5 mJy where a small correction factor of 1.1 must be applied. 
For fluxes greater than 4.5 mJy,  there is no need 
to apply a correction.  

\setcounter{table}{1}
\begin{table}
  \caption{90 cm COSMOS simulation results  }
  \label{tab:simulation}

\begin{tabular}{cccc} 

\hline
Flux bin & Input & Detected & C  ($\sigma_{C}$)\\ 
(mJy)      &          &                &    \\
\hline
~2.00 - ~3.00 & 884 & 398 & 2.2 (0.05)\\
~3.00 - ~4.50 & 534 & 492 & 1.1 (0.07)\\ 
~4.50 - ~6.75 & 552 & 543 & 1.0 (0.09)\\ 
~6.75 - 10.13 & 249 & 258 & 1.0 (0.10)\\
10.13 - 15.19 & 189 & 190 & 1.0 (0.11)\\
15.19 - 22.78 & 206 & 213 & 1.0 (0.12)\\
22.78 - 34.17 & 183 & 183 & 1.0 (0.11)\\
34.17 - 51.26 & 183 & 180 & 1.0 (0.12)\\
51.26 - 76.89 & 169 & 170 & 1.0 (0.10)\\
76.89 - 115.3 & ~62 & ~62 & 1.0 (0.15)\\
 $>$ 115.3     & ~36 & ~36 & 1.0 (0.18)\\
 
 \hline

\end{tabular}

\end{table}

\begin{figure}
\includegraphics[width=\columnwidth]{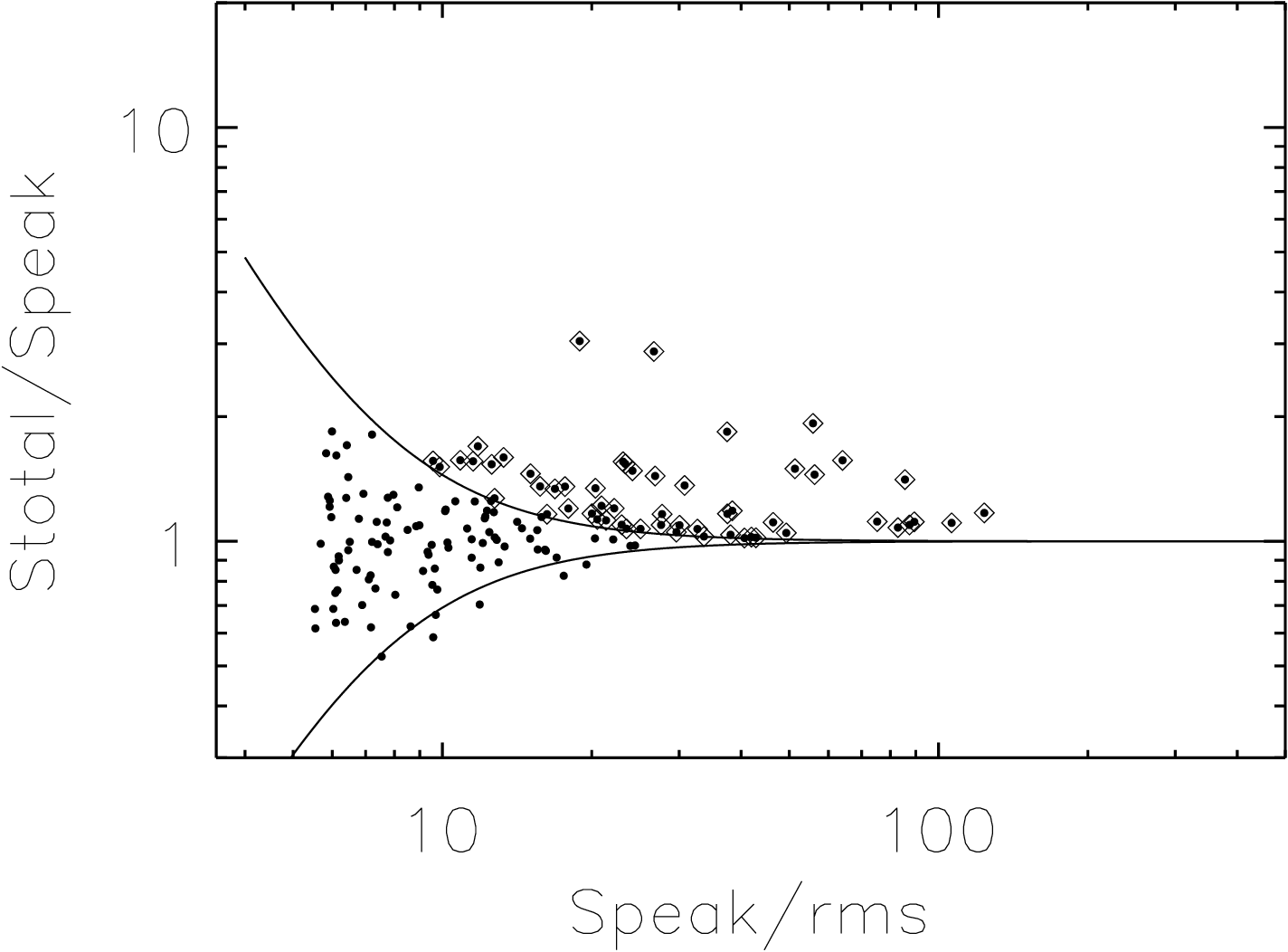}
\caption{  Ratio of the total flux S$_T$ to the peak flux S$_P$ as a function of the
signal-to-noise ratio of the peak flux and the local rms for the 152 single component sources. The solid lines show the
envelopes of the flux ratio distribution used to define resolved sources, indicated by open symbols in the panel.
 }
  \label{fig:res_unres}
\end{figure}

\begin{figure}
\includegraphics[width=\columnwidth]{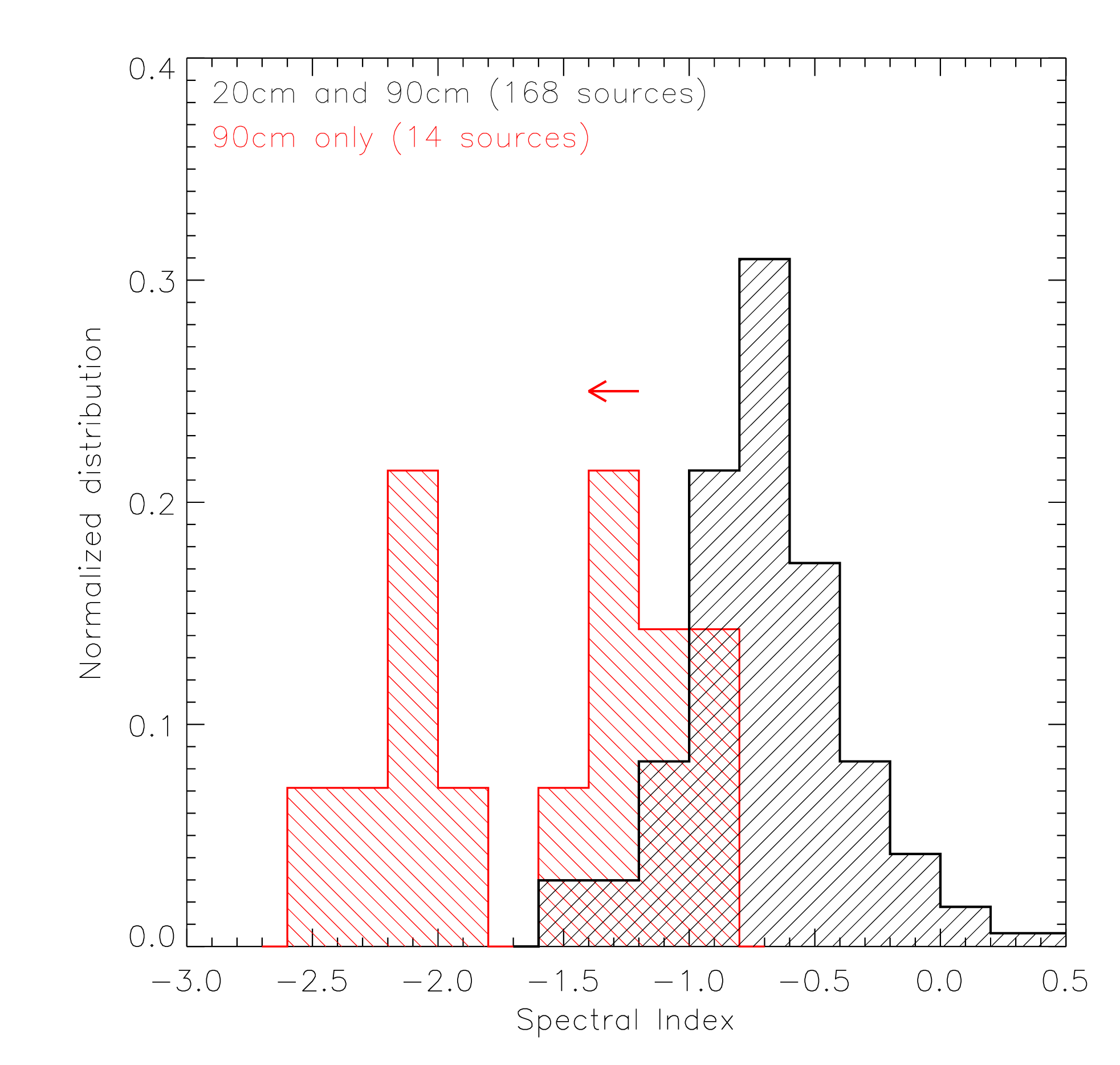}
\caption{ Distribution of 90-20~cm spectral indices for sources detected at both wavelengths (black), and those detected only at 90~cm (red). The latter yield upper limits to the spectral index (indicated by the arrow). }
  \label{fig:alpha}
\end{figure}

\begin{figure}
\includegraphics[width=\columnwidth]{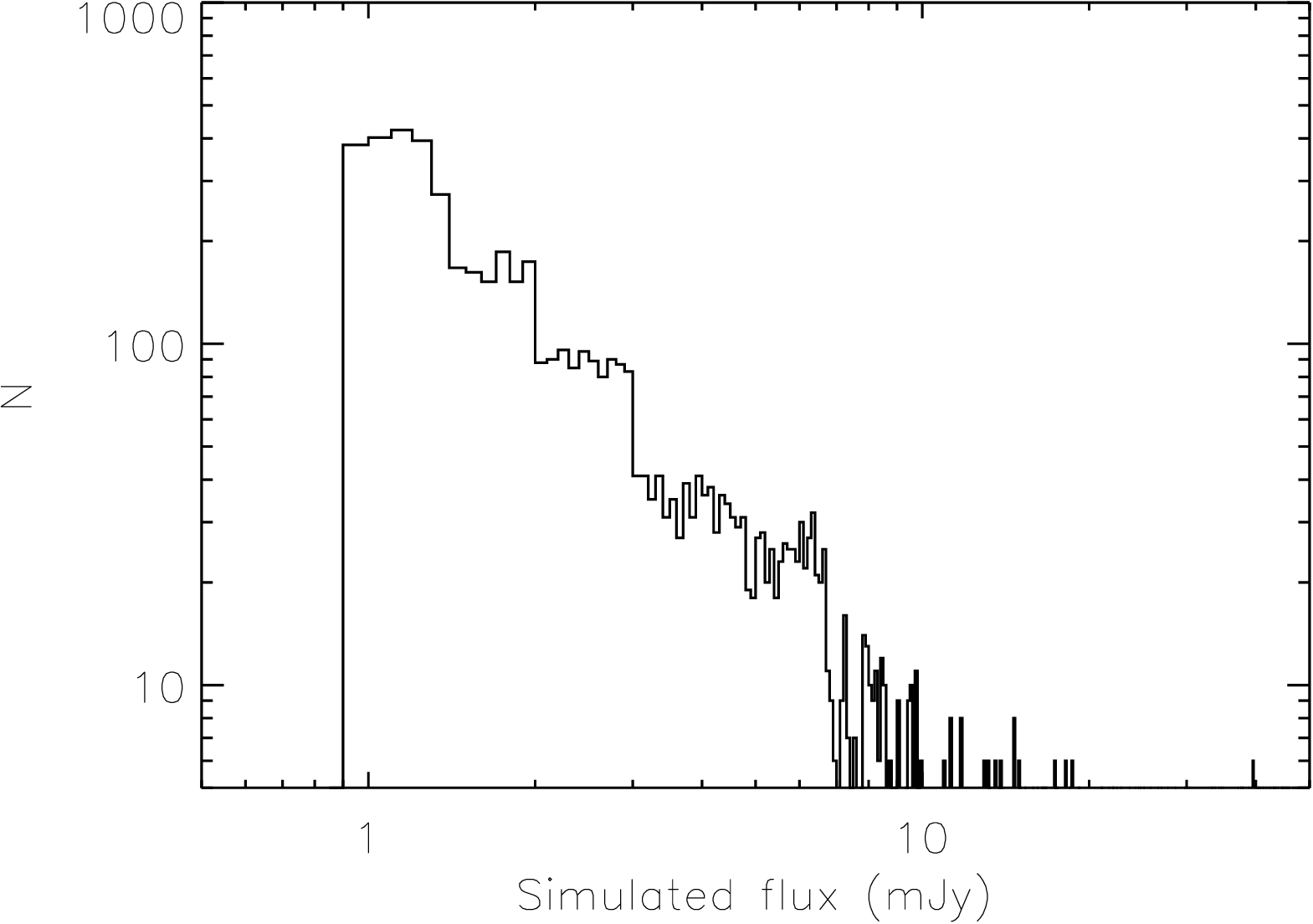}
\caption{ Flux distribution of the 6115 simulated sources used to test the completeness of our 90 cm sample} 
  \label{fig:simflux}
\end{figure}

\section{324~MHz Source Counts} 
\label{sec:counts}

\begin{figure}
\includegraphics[bb=84 390 526 762,width=\columnwidth]{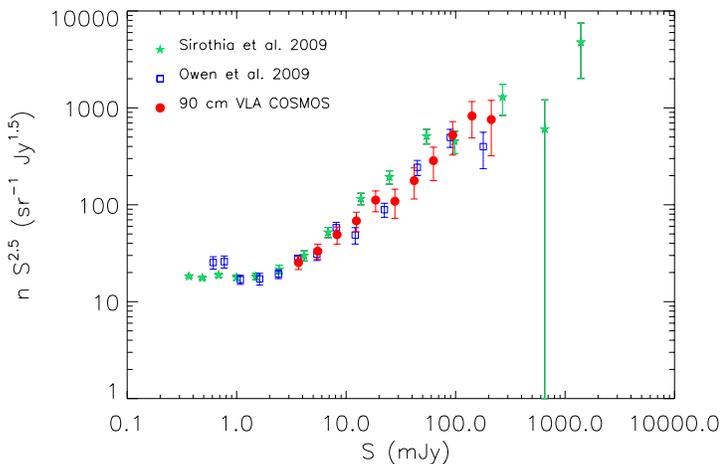}
\caption{ The 90~cm normalized differential source counts for our 90~cm VLA-COSMOS sources (red dots). Source counts at 90 cm from other surveys are also shown:  N1 ELAIS field (\citealt{sirothia09}; filled green stars), $Spitzer$ deep field 1046+59 (\citealt{owen09}; open blue square) }
  \label{fig:counts}
\end{figure}

In order to reduce  problems with possible spurious sources near the flux limit and
effects of incompleteness (see   \t{tab:simulation}  )  we constructed the 
90 cm radio sources counts considering only the 171 sources with a flux density greater than 3.0 mJy, 
corresponding to a S/N $\gtrsim$ 6.0.  
The 90 cm source counts are summarized in  \t{tab:counts} , ~ where, for each flux density bin,  we report 
the mean flux density,  the observed number of sources not corrected for the correction factor C,  the differential source density $dN/dS$
(in sr$^{-1}$ Jy$^{-1}$), the normalized differential counts $nS^{2.5}$  (in sr$^{-1}$ Jy$^{1.5}$) with the estimated 
Poisson error (as $n^{1/2}S^{2.5}$) and the  integrated counts $N(>S)$ (in  deg$^{-2}$). 

\begin{table*}
  \caption{The 90 cm Radio Source Counts for the COSMOS survey}
  \label{tab:counts}

\begin{tabular}{crccrc} 

\hline 
$S$                       &  $<S>$  &  N       &         $dN/dS$                        & $nS^{2.5}$                            & $N(>S)$       \\
(mJy)                      & (mJy)     &            & sr$^{-1}$ Jy$^{-1}$               & sr$^{-1}$ Jy$^{1.5}$             & deg$^{-2}$  \\
\hline
 
    3.00    -~4.50    &   3.67    &    35    &     2.82 $\times$10$^{7}$      & ~~23.04$\pm$     ~~3.91     &  56.91 $\pm$     4.36 \\
    4.50   - ~6.75    &   5.51    &    31    &     1.47 $\times$10$^{7}$      & ~~32.24$\pm$     ~~5.97    &   44.05 $\pm$     3.78 \\
    6.75   -  10.13   &   8.27    &    25    &     7.91 $\times$10$^{6}$      & ~~49.17$\pm$     ~~9.83    &   33.94 $\pm$     3.31 \\
  10.13   -  15.19   &  12.40   &    19    &     3.99 $\times$10$^{6}$      & ~~68.25$\pm$     ~15.66    &   25.81 $\pm$     2.88 \\ 
  15.19   -  22.78   & 18.60    &    17    &     2.37 $\times$10$^{6}$      & ~111.83$\pm$     ~27.12    &   19.66 $\pm$     2.52 \\
  22.78   -  34.17   & 27.90    &      9    &     8.36 $\times$10$^{5}$      & ~108.73$\pm$     ~36.34    &   14.18 $\pm$     2.14 \\
  34.17   -  51.26   & 41.85    &      8    &     4.95 $\times$10$^{5}$      & ~177.52$\pm$     ~62.76    &   11.28 $\pm$     1.91 \\
  51.26    - 76.89   &   62.78  &      7    &     2.89 $\times$10$^{5}$      & ~285.36$\pm$     107.85    &   ~8.70 $\pm$     1.68 \\
  76.89    -115.33  &  94.17   &      7    &     1.93 $\times$10$^{5}$      & ~524.18$\pm$     198.12    &   ~6.44 $\pm$     1.44 \\
 115.33   -173.00  & 141.25  &      6    &     1.10 $\times$10$^{5}$      & ~825.42$\pm$     336.98    &   ~4.19 $\pm$     1.16 \\
 173.00   -259.49  & 211.88  &      3    &     3.67 $\times$10$^{4}$      & ~758.19$\pm$     437.74    &   ~2.26 $\pm$     0.85 \\
 
\hline
         
\end{tabular}
\end{table*}

The normalized differential counts $nS^{2.5}$ multiplied by the correction factor (C) reported in \t{tab:simulation} , are plotted in 
\f{fig:counts}  ~ where, for comparison, the differential source counts obtained from other 90 cm radio surveys are also plotted. 
As shown in \f{fig:counts}  ~ our counts are in very good agreement with previous surveys over the whole flux range  sampled by our 
data ($\sim$ 3 - 250 mJy).  Unfortunately our 90 cm survey is not deep enough to confirm the flattening of the counts in the region below 3 mJy seen in other 90 cm surveys.

\section{Source properties}

\label{sec:multi}

In this section we analyze the redshift distribution, radio luminosities, and radio spectral indices 
of our 90~cm VLA-COSMOS sources.
For this we use the most recent COSMOS photometric redshift catalog (version 1.8 with UltraVISTA data added where applicable; \citealt{ilbert09, ilbert10, mccracken12}), the photometric redshift catalog for XMM-COSMOS sources presented by \citet{salvato09}, and the most recent spectroscopic redshift catalog, comprising a compilation of all spectroscopic observations (public and internal) obtained to-date (zCOSMOS, \citealt{lilly07,lilly09}, 2009; IMACS, \citealt{trump07}; MMT, \citealt{prescott06}; VUDS, \citealt{lefevre14}.; Subaru/FOCAS, Nagao et al., priv. comm.; SDSS DR8, \cite{aihara2011}).

\subsection{Redshift distribution and radio luminosities}

We  restrict our analysis to the area covered by the COSMOS 2  square degree survey. In the 324~MHz map we have identified a total of 182 sources (S/N~$\geq5.5$) over the 3.14  square degree area. Out of these 131 reside within the COSMOS 2 square degree area. Only 6 out of the 131 sources do not have a 20~cm VLA-COSMOS counterpart. 
Given that the 20~cm resolution (and thus the positional accuracy) is by about a factor of 5 better than that in our 90~cm map we hereafter use the 20~cm positions to match the 125 sources detected at 90~cm with other multi-wavelength catalogs. Following \citet{sargent10} we use a  radius of $0.6''$ for the cross-correlation with optical and photometric redshift catalogs (see also Fig.~2 in \citealt{smo08}).  Optical counterparts are identified for 115 out of 125 sources (i.e. 92\%), however 4/115 are located in masked regions and thus no reliable photometry and photometric redshift could be determined for those \citep{ilbert10}. Matching the catalog with the full COSMOS spectroscopic catalog  we identify 49 sources with a reliable spectroscopic redshift.

The cross-correlation of the optical sources with the XMM-COSMOS catalog has been done by \citet{brusa07}. We identify 32 XMM sources in our sample of 115 90~cm sources with 20~cm radio and optical counterparts. 22 of these have reliable spectroscopic redshifts, and for the remainder we use photometric redshifts calculated by \citet{salvato09}.

Hereafter we use spectroscopic redshifts where available, and photometric redshifts otherwise.  The  redshift distribution for the  111 sources detected at 90~cm is shown in \f{fig:photz} . The redshift distribution peaks at $z\sim1$, and shows an extended tail up to $z\sim3$.

In \f{fig:limits} \ we show the 90~cm luminosity distribution of our sources as a function of redshift. We also indicate the 5.5$\sigma$ flux limit (assuming an average spectral index of -0.70), as well as the luminosity limits for various star forming (Milky Way, LIRG, ULIRG, HyLIRG assuming star formation rates of 4, 10, 100, 1000~\msolyr , respectively) and AGN galaxies. For the latter we take a separation of L$_{90~cm}=3\times10^{25}$~\wh \ between high- and low-power radio AGN (e.g. \citealt{kauffmann08}). From the plot it is immediately obvious that the majority of our 90~cm sources are AGN. This is expected based on previous findings in radio surveys, and modeling of radio source populations \citep[e.g][]{wilman08,ballantyne09}.

\subsection{Spectral index as a function of redshift}

The observed spectral index of radio sources has been shown to  decrease as a function of redshift for sources with 1.4~GHz flux densities higher than 10 mJy (e.g.\ \citealt{debreuck00}). An explanation for this is related to a combination of the K-correction of a typically concave-shaped radio synchrotron spectrum ($\log{\mathrm{F}}_{\nu}\propto\alpha\log{\nu}$ where $\alpha$ changes with frequency; see \citealt{miley08}) and an increasing spectral curvature at high redshifts (mainly due to stronger inverse Compton losses in the denser cosmic microwave background at high redshifts; \citealt{krolik91}). However, physical effects such as e.g.\ higher ambient density at higher redshift has to be invoked to explain the effect (see Miley \& de Breuck 2008 for review). Thus, it has been suggested that high spectral indices of radio sources are an efficient proxy for identifying sources at high-redshifts.  Although various efficient UV-NIR drop-out techniques have been identified to-date (e.g.\ \citealt{steidel96}), a purely radio-based high-redshift identifier may prove useful for currently conducted and planned future radio surveys (e.g.\ LOFAR all sky survey, WODAN and EMU) as these will have to rely on radio data only before the availability of next-generation UV-NIR sky surveys (e.g.  LSST, Euclid).

To test whether a trend of steepening of spectral indices  with increasing redshift  exists or not
we show the spectral index as a function of  redshift for our sources detected at 90 and 20~cm with optical counterparts in \f{fig:specindx} . We find no clear evidence of spectral index steepening with redshift in our sample with a flux limit of $\sim2.75$~mJy.  This is consistent with the findings presented by \citet{owen09} who observed the Deep SWIRE field at 324.5~MHz down to an rms of 70~$\mu$Jy/beam, and do not find a large population of ultra-steep spectrum sources down to this limit. 
 Only one source in our sample has a spectral index steeper than $\alpha=-1.3$, the division usually used to identify high redshift source (e.g.\ \citealt{debreuck00}). Nevertheless, our data suggest that a selection of $\alpha<-1$ would  select sources with $z\gtrsim1$.

In the full 90~cm VLA-COSMOS catalog over 3.14 square degrees 182 sources were identified. Out of these 14 do not have counterparts at 20~cm (either in the 20~cm VLA-COSMOS or the FIRST surveys). Out of these 182 sources, in the previous sections we have analyzed 111 sources with 20~cm counterparts and reliable redshifts. Thus, 57 sources with 20~cm counterparts (=182-111-14) remain so-far unaccounted for. In  \f{fig:alpha2} \ we show the spectral index distribution of the full 90~cm sample with 20~cm (VLA-COSMOS or FIRST) counterparts, as well as that of the 57 remaining sources. The distribution of the latter is very similar to that of the first (we find median spectral indices of -0.70 and -0.74 for the full sample, and the subsample, respectively). This suggests similar physical properties of the sources in the two samples. Furthermore, applying a cut of $\alpha<-1$ to the subsample of 57 sources suggests that 7 of these can be identified without other information to be at high redshift, i.e.  $z\gtrsim1$.


\begin{figure}
\includegraphics[bb=50 180 352 432, width=\columnwidth]{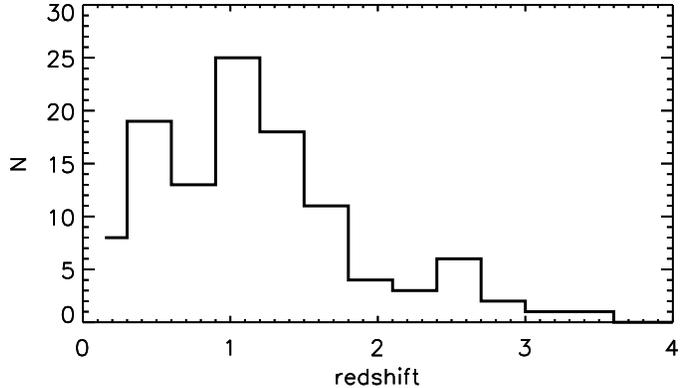}
\caption{ Redshift distribution of the VLA-COSMOS 90~cm sources with NUV-NIR counterparts, and reliable photometric or spectroscopic redshifts. } 
  \label{fig:photz}
\end{figure}

\begin{figure}
\includegraphics[bb=20 0 412 432, width=\columnwidth]{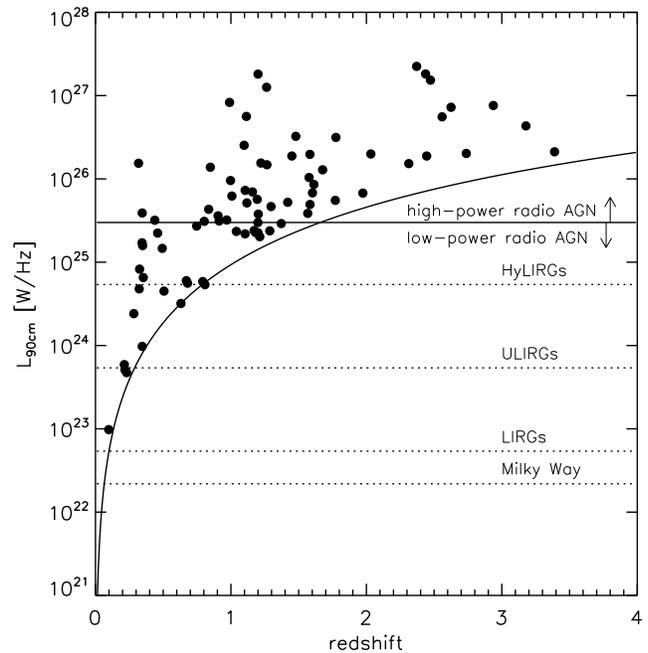}
\caption{ 
90~cm luminosity as a function of redshift for our 324~MHz detected sources.
 The curve shows the 5.5$\sigma$ limit of our 90~cm data (assuming a spectral index of 0.7), and the horizontal lines indicate the locations in this diagram of various types of galaxies (labeled in the panel). Note that at the 90~cm flux limit we predominantly identify AGN galaxies.    }
  \label{fig:limits}
\end{figure}

\begin{figure}
\includegraphics[bb=50 0 412 432, width=\columnwidth]{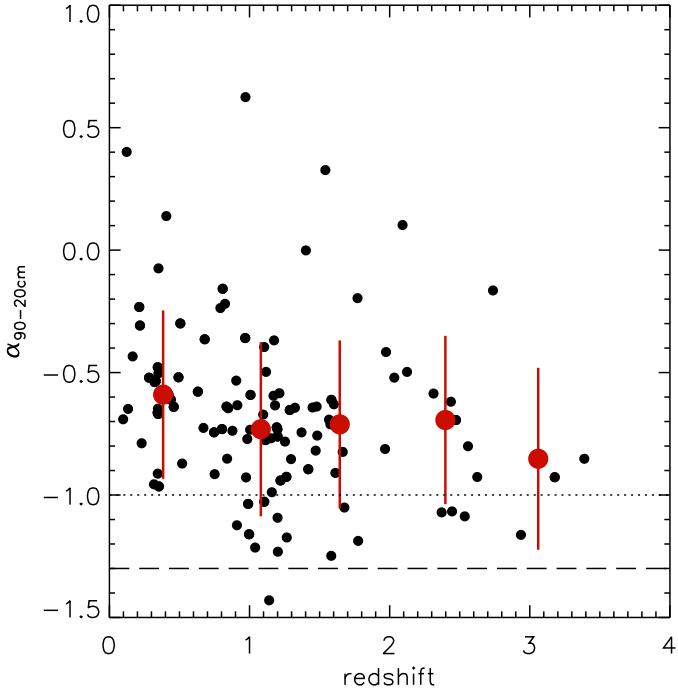}
\caption{ 90-20~cm spectral index vs.\  redshift (spectroscopic where available, otherwise photometric) for our sources detected at 90~cm. The median and root-mean-square scatter of the distribution are also shown (filled dots and vertical error-bars, respectively). The horizontal lines indicate spectral indices of -1 and -1.3.
  \label{fig:specindx}}
\end{figure}

\begin{figure}
\includegraphics[bb=50 0 412 432, width=\columnwidth]{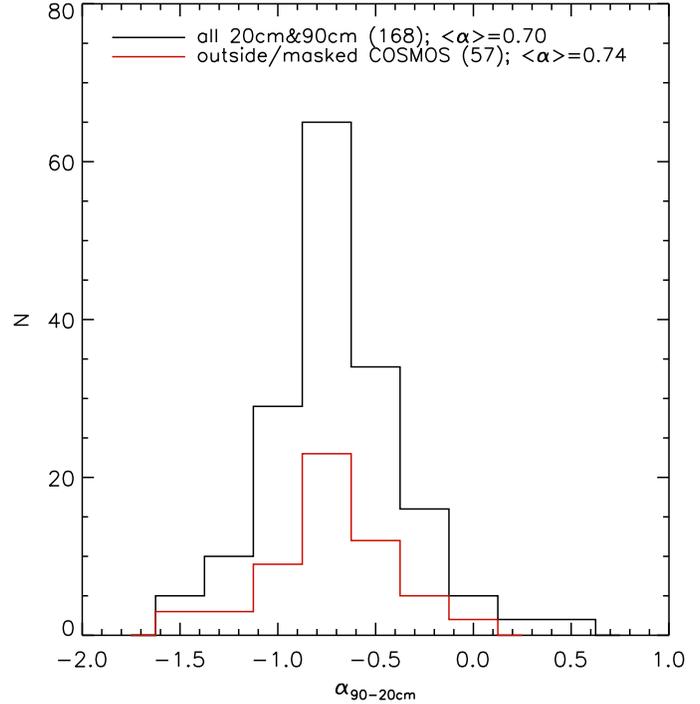}
\caption{ 90-20~cm spectral index distribution for all sources in the 90~cm VLA-COSMOS catalog with 20~cm (VLA-COSMOS Deep or FIRST) counterparts (black line), and the subsample of sources either outside the COSMOS 2sq.deg. area or those within the COSMOS 2 square degree area, but without optical counterparts or reliable redshifts (red line).  The similar spectral index distribution suggests similar physical properties of the two samples. }
  \label{fig:alpha2}
\end{figure}

\section{Summary}
\label{sec:summary}

We have presented a 90~cm VLA map of the COSMOS field, comprising a circular area of 3.14 square degrees at $6.0''\times5.3''$ angular resolution with an average rms of 0.5~mJy/beam. The extracted catalog contains 182 sources, 30 of which consist of multiple components. Using Monte Carlo artificial source simulations we have derived the completeness correction for the catalog. Our source counts agree very well with those from previous studies, verifying the validity of the catalog.
As expected based on previous findings in radio surveys, and modeling of radio source populations \citep[e.g][]{wilman08,ballantyne09} our sample is AGN dominated, as inferred based on 90~cm radio luminosity considerations.

Combining the 90~cm with our 20~cm VLA-COSMOS data we infer a median 90-20~cm spectral index of -0.70, with an interquartile range of -0.90 to -0.53. This is consistent with values found at 90 vs.\ 20~cm wavelengths \citep{owen09}, as well as that typically found for radio sources at high (GHz) radio frequencies (e.g.\ Kimball \& Ivezi\'{c} 2008).

Making use of the COSMOS photometric and spectroscopic redshifts  of our 90~cm sources we find no strong evidence for steepening of the spectral index with redshift. 
The sources in our sample with spectral indices steeper than -1 all lie at $z\gtrsim1$, in agreement with the idea that ultra-steep-spectrum radio sources may trace intermediate-redshift galaxies (z $\gtrsim$ 1).

\section*{acknowledgement}
VS acknowledges the European Unions Seventh Frame-work program grant agreement 333654 (CIG, 'AGN feedback') and the Australian Group of Eight European Fellowship 2013. AK acknowledges support by the Collaborative Research Council 956, sub-project A1, funded by the Deutsche Forschungsgemeinschaft (DFG). The National Radio Astronomy Observatory is a facility of the National Science Foundation operated under cooperative agreement by Associated Universities, Inc.
 
\bibliographystyle{mn2e}
\bibliography{reflistCOSMOS}

\begin{thebibliography}{57}
\expandafter\ifx\csname natexlab\endcsname\relax\def\natexlab#1{#1}\fi

\bibitem[{{Aihara} {et~al}\mbox{.}(2011){Aihara}, {Allende Prieto}, {An},
  {Anderson}, {Aubourg}, {Balbinot}, {Beers}, {Berlind}, {Bickerton},
  {Bizyaev}, {Blanton}, {Bochanski}, {Bolton}, {Bovy}, {Brandt}, {Brinkmann},
  {Brown}, {Brownstein}, {Busca}, {Campbell}, {Carr}, {Chen}, {Chiappini},
  {Comparat}, {Connolly}, {Cortes}, {Croft}, {Cuesta}, {da Costa}, {Davenport},
  {Dawson}, {Dhital}, {Ealet}, {Ebelke}, {Edmondson}, {Eisenstein},
  {Escoffier}, {Esposito}, {Evans}, {Fan}, {Femen{\'{\i}}a Castell{\'a}},
  {Font-Ribera}, {Frinchaboy}, {Ge}, {Gillespie}, {Gilmore}, {Gonz{\'a}lez
  Hern{\'a}ndez}, {Gott}, {Gould}, {Grebel}, {Gunn}, {Hamilton}, {Harding},
  {Harris}, {Hawley}, {Hearty}, {Ho}, {Hogg}, {Holtzman}, {Honscheid}, {Inada},
  {Ivans}, {Jiang}, {Johnson}, {Jordan}, {Jordan}, {Kazin}, {Kirkby}, {Klaene},
  {Knapp}, {Kneib}, {Kochanek}, {Koesterke}, {Kollmeier}, {Kron}, {Lampeitl},
  {Lang}, {Le Goff}, {Lee}, {Lin}, {Long}, {Loomis}, {Lucatello}, {Lundgren},
  {Lupton}, {Ma}, {MacDonald}, {Mahadevan}, {Maia}, {Makler}, {Malanushenko},
  {Malanushenko}, {Mandelbaum}, {Maraston}, {Margala}, {Masters}, {McBride},
  {McGehee}, {McGreer}, {M{\'e}nard}, {Miralda-Escud{\'e}}, {Morrison},
  {Mullally}, {Muna}, {Munn}, {Murayama}, {Myers}, {Naugle}, {Neto}, {Nguyen},
  {Nichol}, {O'Connell}, {Ogando}, {Olmstead}, {Oravetz}, {Padmanabhan},
  {Palanque-Delabrouille}, {Pan}, {Pandey}, {P{\^a}ris}, {Percival},
  {Petitjean}, {Pfaffenberger}, {Pforr}, {Phleps}, {Pichon}, {Pieri}, {Prada},
  {Price-Whelan}, {Raddick}, {Ramos}, {Reyl{\'e}}, {Rich}, {Richards}, {Rix},
  {Robin}, {Rocha-Pinto}, {Rockosi}, {Roe}, {Rollinde}, {Ross}, {Ross},
  {Rossetto}, {S{\'a}nchez}, {Sayres}, {Schlegel}, {Schlesinger}, {Schmidt},
  {Schneider}, {Sheldon}, {Shu}, {Simmerer}, {Simmons}, {Sivarani}, {Snedden},
  {Sobeck}, {Steinmetz}, {Strauss}, {Szalay}, {Tanaka}, {Thakar}, {Thomas},
  {Tinker}, {Tofflemire}, {Tojeiro}, {Tremonti}, {Vandenberg}, {Vargas
  Maga{\~n}a}, {Verde}, {Vogt}, {Wake}, {Wang}, {Weaver}, {Weinberg}, {White},
  {White}, {Yanny}, {Yasuda}, {Yeche}, \& {Zehavi}}]{aihara2011}
{Aihara} H. {et~al.}, 2011, \apjs, 193, 29

\bibitem[{{Ballantyne}(2009)}]{ballantyne09}
{Ballantyne} D.~R., 2009, \apj, 698, 1033

\bibitem[{{Best} \& {Heckman}(2012)}]{best12}
{Best} P.~N., {Heckman} T.~M., 2012, \mnras, 421, 1569

\bibitem[{{B{\^i}rzan} {et~al}\mbox{.}(2008){B{\^i}rzan}, {McNamara}, {Nulsen},
  {Carilli}, \& {Wise}}]{birzan08}
{B{\^i}rzan} L., {McNamara} B.~R., {Nulsen} P.~E.~J., {Carilli} C.~L., {Wise}
  M.~W., 2008, \apj, 686, 859

\bibitem[{{B{\^i}rzan} {et~al}\mbox{.}(2004){B{\^i}rzan}, {Rafferty},
  {McNamara}, {Wise}, \& {Nulsen}}]{birzan04}
{B{\^i}rzan} L., {Rafferty} D.~A., {McNamara} B.~R., {Wise} M.~W., {Nulsen}
  P.~E.~J., 2004, \apj, 607, 800

\bibitem[{{Bondi} {et~al}\mbox{.}(2007){Bondi}, {Ciliegi}, {Venturi},
  {Dallacasa}, {Bardelli}, {Zucca}, {Athreya}, {Gregorini}, {Zanichelli}, {Le
  F{\`e}vre}, {Contini}, {Garilli}, {Iovino}, {Temporin}, \&
  {Vergani}}]{bondi07}
{Bondi} M. {et~al.}, 2007, \aap, 463, 519

\bibitem[{{Bondi} {et~al}\mbox{.}(2003){Bondi}, {Ciliegi}, {Zamorani},
  {Gregorini}, {Vettolani}, {Parma}, {de Ruiter}, {Le Fevre}, {Arnaboldi},
  {Guzzo}, {Maccagni}, {Scaramella}, {Adami}, {Bardelli}, {Bolzonella},
  {Bottini}, {Cappi}, {Foucaud}, {Franzetti}, {Garilli}, {Gwyn}, {Ilbert},
  {Iovino}, {Le Brun}, {Marano}, {Marinoni}, {McCracken}, {Meneux}, {Pollo},
  {Pozzetti}, {Radovich}, {Ripepi}, {Rizzo}, {Scodeggio}, {Tresse},
  {Zanichelli}, \& {Zucca}}]{bondi03}
{Bondi} M. {et~al.}, 2003, \aap, 403, 857

\bibitem[{{Bower} {et~al}\mbox{.}(2006){Bower}, {Benson}, {Malbon}, {Helly},
  {Frenk}, {Baugh}, {Cole}, \& {Lacey}}]{bower06}
{Bower} R.~G., {Benson} A.~J., {Malbon} R., {Helly} J.~C., {Frenk} C.~S.,
  {Baugh} C.~M., {Cole} S., {Lacey} C.~G., 2006, \mnras, 370, 645

\bibitem[{{Brusa} {et~al}\mbox{.}(2007){Brusa}, {Zamorani}, {Comastri},
  {Hasinger}, {Cappelluti}, {Civano}, {Finoguenov}, {Mainieri}, {Salvato},
  {Vignali}, {Elvis}, {Fiore}, {Gilli}, {Impey}, {Lilly}, {Mignoli},
  {Silverman}, {Trump}, {Urry}, {Bender}, {Capak}, {Huchra}, {Kneib},
  {Koekemoer}, {Leauthaud}, {Lehmann}, {Massey}, {Matute}, {McCarthy},
  {McCracken}, {Rhodes}, {Scoville}, {Taniguchi}, \& {Thompson}}]{brusa07}
{Brusa} M. {et~al.}, 2007, \apjs, 172, 353

\bibitem[{{Capak} {et~al}\mbox{.}(2007){Capak}, {Aussel}, {Ajiki}, {McCracken},
  {Mobasher}, {Scoville}, {Shopbell}, {Taniguchi}, {Thompson}, {Tribiano},
  {Sasaki}, {Blain}, {Brusa}, {Carilli}, {Comastri}, {Carollo}, {Cassata},
  {Colbert}, {Ellis}, {Elvis}, {Giavalisco}, {Green}, {Guzzo}, {Hasinger},
  {Ilbert}, {Impey}, {Jahnke}, {Kartaltepe}, {Kneib}, {Koda}, {Koekemoer},
  {Komiyama}, {Leauthaud}, {Le Fevre}, {Lilly}, {Liu}, {Massey}, {Miyazaki},
  {Murayama}, {Nagao}, {Peacock}, {Pickles}, {Porciani}, {Renzini}, {Rhodes},
  {Rich}, {Salvato}, {Sanders}, {Scarlata}, {Schiminovich}, {Schinnerer},
  {Scodeggio}, {Sheth}, {Shioya}, {Tasca}, {Taylor}, {Yan}, \&
  {Zamorani}}]{capak07}
{Capak} P. {et~al.}, 2007, \apjs, 172, 99

\bibitem[{{Condon}(1997)}]{condon97}
{Condon} J.~J., 1997, PASP, 109, 166

\bibitem[{{Croton} {et~al}\mbox{.}(2006){Croton}, {Springel}, {White}, {De
  Lucia}, {Frenk}, {Gao}, {Jenkins}, {Kauffmann}, {Navarro}, \&
  {Yoshida}}]{croton06}
{Croton} D.~J. {et~al.}, 2006, \mnras, 365, 11

\bibitem[{{De Breuck} {et~al}\mbox{.}(2002){De Breuck}, {Tang}, {de Bruyn},
  {R{\"o}ttgering}, \& {van Breugel}}]{carlos02}
{De Breuck} C., {Tang} Y., {de Bruyn} A.~G., {R{\"o}ttgering} H., {van Breugel}
  W., 2002, \aap, 394, 59

\bibitem[{{De Breuck} {et~al}\mbox{.}(2000){De Breuck}, {van Breugel},
  {R{\"o}ttgering}, \& {Miley}}]{debreuck00}
{De Breuck} C., {van Breugel} W., {R{\"o}ttgering} H.~J.~A., {Miley} G., 2000,
  \aaps, 143, 303

\bibitem[{{Elvis} {et~al}\mbox{.}(2009){Elvis}, {Civano}, {Vignali},
  {Puccetti}, {Fiore}, {Cappelluti}, {Aldcroft}, {Fruscione}, {Zamorani},
  {Comastri}, {Brusa}, {Gilli}, {Miyaji}, {Damiani}, {Koekemoer}, {Finoguenov},
  {Brunner}, {Urry}, {Silverman}, {Mainieri}, {Hasinger}, {Griffiths},
  {Carollo}, {Hao}, {Guzzo}, {Blain}, {Calzetti}, {Carilli}, {Capak}, {Ettori},
  {Fabbiano}, {Impey}, {Lilly}, {Mobasher}, {Rich}, {Salvato}, {Sanders},
  {Schinnerer}, {Scoville}, {Shopbell}, {Taylor}, {Taniguchi}, \&
  {Volonteri}}]{elvis09}
{Elvis} M. {et~al.}, 2009, \apjs, 184, 158

\bibitem[{{Evans} {et~al}\mbox{.}(2006){Evans}, {Worrall}, {Hardcastle},
  {Kraft}, \& {Birkinshaw}}]{evans06}
{Evans} D.~A., {Worrall} D.~M., {Hardcastle} M.~J., {Kraft} R.~P., {Birkinshaw}
  M., 2006, \apj, 642, 96

\bibitem[{{Greisen}(1990)}]{greisen90}
{Greisen} E.~W., 1990, in Acquisition, Processing and Archiving of Astronomical
  Images, {Longo} G., {Sedmak} G., eds., pp. 125--142

\bibitem[{{Hardcastle}, {Evans} \& {Croston}(2007){Hardcastle}, {Evans}, \&
  {Croston}}]{hardcastle07}
{Hardcastle} M.~J., {Evans} D.~A., {Croston} J.~H., 2007, \mnras, 376, 1849

\bibitem[{{Hasinger} {et~al}\mbox{.}(2007){Hasinger}, {Cappelluti}, {Brunner},
  {Brusa}, {Comastri}, {Elvis}, {Finoguenov}, {Fiore}, {Franceschini}, {Gilli},
  {Griffiths}, {Lehmann}, {Mainieri}, {Matt}, {Matute}, {Miyaji}, {Molendi},
  {Paltani}, {Sanders}, {Scoville}, {Tresse}, {Urry}, {Vettolani}, \&
  {Zamorani}}]{hasinger07}
{Hasinger} G. {et~al.}, 2007, \apjs, 172, 29

\bibitem[{{Ho}(2005)}]{ho05}
{Ho} L.~C., 2005, \apss, 300, 219

\bibitem[{{Hopkins} {et~al}\mbox{.}(2003){Hopkins}, {Afonso}, {Chan}, {Cram},
  {Georgakakis}, \& {Mobasher}}]{hopkins03}
{Hopkins} A.~M., {Afonso} J., {Chan} B., {Cram} L.~E., {Georgakakis} A.,
  {Mobasher} B., 2003, \aj, 125, 465

\bibitem[{{Ilbert} {et~al}\mbox{.}(2009){Ilbert}, {Capak}, {Salvato}, {Aussel},
  {McCracken}, {Sanders}, {Scoville}, {Kartaltepe}, {Arnouts}, {Le Floc'h},
  {Mobasher}, {Taniguchi}, {Lamareille}, {Leauthaud}, {Sasaki}, {Thompson},
  {Zamojski}, {Zamorani}, {Bardelli}, {Bolzonella}, {Bongiorno}, {Brusa},
  {Caputi}, {Carollo}, {Contini}, {Cook}, {Coppa}, {Cucciati}, {de la Torre},
  {de Ravel}, {Franzetti}, {Garilli}, {Hasinger}, {Iovino}, {Kampczyk},
  {Kneib}, {Knobel}, {Kovac}, {Le Borgne}, {Le Brun}, {F{\`e}vre}, {Lilly},
  {Looper}, {Maier}, {Mainieri}, {Mellier}, {Mignoli}, {Murayama}, {Pell{\`o}},
  {Peng}, {P{\'e}rez-Montero}, {Renzini}, {Ricciardelli}, {Schiminovich},
  {Scodeggio}, {Shioya}, {Silverman}, {Surace}, {Tanaka}, {Tasca}, {Tresse},
  {Vergani}, \& {Zucca}}]{ilbert09}
{Ilbert} O. {et~al.}, 2009, \apj, 690, 1236

\bibitem[{{Ilbert} {et~al}\mbox{.}(2010){Ilbert}, {Salvato}, {Le Floc'h},
  {Aussel}, {Capak}, {McCracken}, {Mobasher}, {Kartaltepe}, {Scoville},
  {Sanders}, {Arnouts}, {Bundy}, {Cassata}, {Kneib}, {Koekemoer}, {Le
  F{\`e}vre}, {Lilly}, {Surace}, {Taniguchi}, {Tasca}, {Thompson}, {Tresse},
  {Zamojski}, {Zamorani}, \& {Zucca}}]{ilbert10}
{Ilbert} O. {et~al.}, 2010, \apj, 709, 644

\bibitem[{{Jeli{\'c}} {et~al}\mbox{.}(2012){Jeli{\'c}}, {Smol{\v c}i{\'c}},
  {Finoguenov}, {Tanaka}, {Civano}, {Schinnerer}, {Cappelluti}, \&
  {Koekemoer}}]{jelic12}
{Jeli{\'c}} V., {Smol{\v c}i{\'c}} V., {Finoguenov} A., {Tanaka} M., {Civano}
  F., {Schinnerer} E., {Cappelluti} N., {Koekemoer} A., 2012, \mnras, 423, 2753

\bibitem[{{Johnston} {et~al}\mbox{.}(2007){Johnston}, {Bailes}, {Bartel},
  {Baugh}, {Bietenholz}, {Blake}, {Braun}, {Brown}, {Chatterjee}, {Darling},
  {Deller}, {Dodson}, {Edwards}, {Ekers}, {Ellingsen}, {Feain}, {Gaensler},
  {Haverkorn}, {Hobbs}, {Hopkins}, {Jackson}, {James}, {Joncas}, {Kaspi},
  {Kilborn}, {Koribalski}, {Kothes}, {Landecker}, {Lenc}, {Lovell}, {Macquart},
  {Manchester}, {Matthews}, {McClure-Griffiths}, {Norris}, {Pen}, {Phillips},
  {Power}, {Protheroe}, {Sadler}, {Schmidt}, {Stairs}, {Staveley-Smith},
  {Stil}, {Taylor}, {Tingay}, {Tzioumis}, {Walker}, {Wall}, \&
  {Wolleben}}]{johnston07}
{Johnston} S. {et~al.}, 2007, Publications of the Astronomical Society of
  Australia, 24, 174

\bibitem[{{Kauffmann}, {Heckman} \& {Best}(2008){Kauffmann}, {Heckman}, \&
  {Best}}]{kauffmann08}
{Kauffmann} G., {Heckman} T.~M., {Best} P.~N., 2008, \mnras, 384, 953

\bibitem[{{Kimball} \& {Ivezi{\'c}}(2008)}]{kimball08}
{Kimball} A.~E., {Ivezi{\'c}} {\v Z}., 2008, \aj, 136, 684

\bibitem[{{Koekemoer} {et~al}\mbox{.}(2007){Koekemoer}, {Aussel}, {Calzetti},
  {Capak}, {Giavalisco}, {Kneib}, {Leauthaud}, {Le F{\`e}vre}, {McCracken},
  {Massey}, {Mobasher}, {Rhodes}, {Scoville}, \& {Shopbell}}]{koekemoer07}
{Koekemoer} A.~M. {et~al.}, 2007, \apjs, 172, 196

\bibitem[{{Krolik} \& {Chen}(1991)}]{krolik91}
{Krolik} J.~H., {Chen} W., 1991, \aj, 102, 1659

\bibitem[{{Le Fevre} {et~al}\mbox{.}(2014){Le Fevre}, {Tasca}, {Cassata},
  {Garilli}, {Le Brun}, {Maccagni}, {Pentericci}, {Thomas}, {Vanzella},
  {Zamorani}, {Zucca}, {Amorin}, {Bardelli}, {Capak}, {Cassara}, {Castellano},
  {Cimatti}, {Cuby}, {Cucciati}, {de la Torre}, {Durkalec}, {Fontana},
  {Giavalisco}, {Grazian}, {Hathi}, {Ilbert}, {Lemaux}, {Moreau}, {Paltani},
  {Ribeiro}, {Salvato}, {Schaerer}, {Scodeggio}, {Sommariva}, {Talia},
  {Taniguchi}, {Tresse}, {Vergani}, {Wang}, {Charlot}, {Contini}, {Fotopoulo},
  {Lopez-Sanjuan}, {Mellier}, \& {Scoville}}]{lefevre14}
{Le Fevre} O. {et~al.}, 2014, ArXiv e-prints

\bibitem[{{Lilly} {et~al}\mbox{.}(2009){Lilly}, {Le Brun}, {Maier}, {Mainieri},
  {Mignoli}, {Scodeggio}, {Zamorani}, {Carollo}, {Contini}, {Kneib}, {Le
  F{\`e}vre}, {Renzini}, {Bardelli}, {Bolzonella}, {Bongiorno}, {Caputi},
  {Coppa}, {Cucciati}, {de la Torre}, {de Ravel}, {Franzetti}, {Garilli},
  {Iovino}, {Kampczyk}, {Kovac}, {Knobel}, {Lamareille}, {Le Borgne}, {Pello},
  {Peng}, {P{\'e}rez-Montero}, {Ricciardelli}, {Silverman}, {Tanaka}, {Tasca},
  {Tresse}, {Vergani}, {Zucca}, {Ilbert}, {Salvato}, {Oesch}, {Abbas},
  {Bottini}, {Capak}, {Cappi}, {Cassata}, {Cimatti}, {Elvis}, {Fumana},
  {Guzzo}, {Hasinger}, {Koekemoer}, {Leauthaud}, {Maccagni}, {Marinoni},
  {McCracken}, {Memeo}, {Meneux}, {Porciani}, {Pozzetti}, {Sanders},
  {Scaramella}, {Scarlata}, {Scoville}, {Shopbell}, \& {Taniguchi}}]{lilly09}
{Lilly} S.~J. {et~al.}, 2009, \apjs, 184, 218

\bibitem[{{Lilly} {et~al}\mbox{.}(2007){Lilly}, {Le F{\`e}vre}, {Renzini},
  {Zamorani}, {Scodeggio}, {Contini}, {Carollo}, {Hasinger}, {Kneib}, {Iovino},
  {Le Brun}, {Maier}, {Mainieri}, {Mignoli}, {Silverman}, {Tasca},
  {Bolzonella}, {Bongiorno}, {Bottini}, {Capak}, {Caputi}, {Cimatti},
  {Cucciati}, {Daddi}, {Feldmann}, {Franzetti}, {Garilli}, {Guzzo}, {Ilbert},
  {Kampczyk}, {Kovac}, {Lamareille}, {Leauthaud}, {Borgne}, {McCracken},
  {Marinoni}, {Pello}, {Ricciardelli}, {Scarlata}, {Vergani}, {Sanders},
  {Schinnerer}, {Scoville}, {Taniguchi}, {Arnouts}, {Aussel}, {Bardelli},
  {Brusa}, {Cappi}, {Ciliegi}, {Finoguenov}, {Foucaud}, {Franceschini},
  {Halliday}, {Impey}, {Knobel}, {Koekemoer}, {Kurk}, {Maccagni}, {Maddox},
  {Marano}, {Marconi}, {Meneux}, {Mobasher}, {Moreau}, {Peacock}, {Porciani},
  {Pozzetti}, {Scaramella}, {Schiminovich}, {Shopbell}, {Smail}, {Thompson},
  {Tresse}, {Vettolani}, {Zanichelli}, \& {Zucca}}]{lilly07}
{Lilly} S.~J. {et~al.}, 2007, \apjs, 172, 70

\bibitem[{{McCracken} {et~al}\mbox{.}(2012){McCracken}, {Milvang-Jensen},
  {Dunlop}, {Franx}, {Fynbo}, {Le F{\`e}vre}, {Holt}, {Caputi}, {Goranova},
  {Buitrago}, {Emerson}, {Freudling}, {Hudelot}, {L{\'o}pez-Sanjuan},
  {Magnard}, {Mellier}, {M{\o}ller}, {Nilsson}, {Sutherland}, {Tasca}, \&
  {Zabl}}]{mccracken12}
{McCracken} H.~J. {et~al.}, 2012, \aap, 544, A156

\bibitem[{{Miley} \& {De Breuck}(2008)}]{miley08}
{Miley} G., {De Breuck} C., 2008, \aapr, 15, 67

\bibitem[{{Mobasher} {et~al}\mbox{.}(2007){Mobasher}, {Capak}, {Scoville},
  {Dahlen}, {Salvato}, {Aussel}, {Thompson}, {Feldmann}, {Tasca}, {Le Fevre},
  {Lilly}, {Carollo}, {Kartaltepe}, {McCracken}, {Mould}, {Renzini}, {Sanders},
  {Shopbell}, {Taniguchi}, {Ajiki}, {Shioya}, {Contini}, {Giavalisco},
  {Ilbert}, {Iovino}, {Le Brun}, {Mainieri}, {Mignoli}, \&
  {Scodeggio}}]{mobasher07}
{Mobasher} B. {et~al.}, 2007, \apjs, 172, 117

\bibitem[{{Norris} {et~al}\mbox{.}(2011){Norris}, {Hopkins}, {Afonso}, {Brown},
  {Condon}, {Dunne}, {Feain}, {Hollow}, {Jarvis}, {Johnston-Hollitt}, {Lenc},
  {Middelberg}, {Padovani}, {Prandoni}, {Rudnick}, {Seymour}, {Umana},
  {Andernach}, {Alexander}, {Appleton}, {Bacon}, {Banfield}, {Becker}, {Brown},
  {Ciliegi}, {Jackson}, {Eales}, {Edge}, {Gaensler}, {Giovannini}, {Hales},
  {Hancock}, {Huynh}, {Ibar}, {Ivison}, {Kennicutt}, {Kimball}, {Koekemoer},
  {Koribalski}, {L{\'o}pez-S{\'a}nchez}, {Mao}, {Murphy}, {Messias},
  {Pimbblet}, {Raccanelli}, {Randall}, {Reiprich}, {Roseboom},
  {R{\"o}ttgering}, {Saikia}, {Sharp}, {Slee}, {Smail}, {Thompson}, {Urquhart},
  {Wall}, \& {Zhao}}]{norris11}
{Norris} R.~P. {et~al.}, 2011, Publications of the Astronomical Society of
  Australia, 28, 215

\bibitem[{{Oklop{\v c}i{\'c}} {et~al}\mbox{.}(2010){Oklop{\v c}i{\'c}},
  {Smol{\v c}i{\'c}}, {Giodini}, {Zamorani}, {B{\^i}rzan}, {Schinnerer},
  {Carilli}, {Finoguenov}, {Lilly}, {Koekemoer}, \& {Scoville}}]{oklopcic10}
{Oklop{\v c}i{\'c}} A. {et~al.}, 2010, \apj, 713, 484

\bibitem[{{Owen} {et~al}\mbox{.}(2009){Owen}, {Morrison}, {Klimek}, \&
  {Greisen}}]{owen09}
{Owen} F.~N., {Morrison} G.~E., {Klimek} M.~D., {Greisen} E.~W., 2009, \aj,
  137, 4846

\bibitem[{{Prescott} {et~al}\mbox{.}(2006){Prescott}, {Impey}, {Cool}, \&
  {Scoville}}]{prescott06}
{Prescott} M.~K.~M., {Impey} C.~D., {Cool} R.~J., {Scoville} N.~Z., 2006, \apj,
  644, 100

\bibitem[{{Salvato} {et~al}\mbox{.}(2009){Salvato}, {Hasinger}, {Ilbert},
  {Zamorani}, {Brusa}, {Scoville}, {Rau}, {Capak}, {Arnouts}, {Aussel},
  {Bolzonella}, {Buongiorno}, {Cappelluti}, {Caputi}, {Civano}, {Cook},
  {Elvis}, {Gilli}, {Jahnke}, {Kartaltepe}, {Impey}, {Lamareille}, {Le Floc'h},
  {Lilly}, {Mainieri}, {McCarthy}, {McCracken}, {Mignoli}, {Mobasher},
  {Murayama}, {Sasaki}, {Sanders}, {Schiminovich}, {Shioya}, {Shopbell},
  {Silverman}, {Smol{\v c}i{\'c}}, {Surace}, {Taniguchi}, {Thompson}, {Trump},
  {Urry}, \& {Zamojski}}]{salvato09}
{Salvato} M. {et~al.}, 2009, \apj, 690, 1250

\bibitem[{{Sanders} {et~al}\mbox{.}(2007){Sanders}, {Salvato}, {Aussel},
  {Ilbert}, {Scoville}, {Surace}, {Frayer}, {Sheth}, {Helou}, {Brooke},
  {Bhattacharya}, {Yan}, {Kartaltepe}, {Barnes}, {Blain}, {Calzetti}, {Capak},
  {Carilli}, {Carollo}, {Comastri}, {Daddi}, {Ellis}, {Elvis}, {Fall},
  {Franceschini}, {Giavalisco}, {Hasinger}, {Impey}, {Koekemoer}, {Le
  F{\`e}vre}, {Lilly}, {Liu}, {McCracken}, {Mobasher}, {Renzini}, {Rich},
  {Schinnerer}, {Shopbell}, {Taniguchi}, {Thompson}, {Urry}, \&
  {Williams}}]{sanders07}
{Sanders} D.~B. {et~al.}, 2007, \apjs, 172, 86

\bibitem[{{Sargent} {et~al}\mbox{.}(2010){Sargent}, {Schinnerer}, {Murphy},
  {Aussel}, {Le Floc'h}, {Frayer}, {Mart{\'{\i}}nez-Sansigre}, {Oesch},
  {Salvato}, {Smol{\v c}i{\'c}}, {Zamorani}, {Brusa}, {Cappelluti}, {Carilli},
  {Carollo}, {Ilbert}, {Kartaltepe}, {Koekemoer}, {Lilly}, {Sanders}, \&
  {Scoville}}]{sargent10}
{Sargent} M.~T. {et~al.}, 2010, \apjs, 186, 341

\bibitem[{{Schinnerer} {et~al}\mbox{.}(2004){Schinnerer}, {Carilli},
  {Scoville}, {Bondi}, {Ciliegi}, {Vettolani}, {Le F{\`e}vre}, {Koekemoer},
  {Bertoldi}, \& {Impey}}]{schinnerer04}
{Schinnerer} E. {et~al.}, 2004, \aj, 128, 1974

\bibitem[{{Schinnerer} {et~al}\mbox{.}(2010){Schinnerer}, {Sargent}, {Bondi},
  {Smol{\v c}i{\'c}}, {Datta}, {Carilli}, {Bertoldi}, {Blain}, {Ciliegi},
  {Koekemoer}, \& {Scoville}}]{schinnerer10}
{Schinnerer} E. {et~al.}, 2010, \apjs, 188, 384

\bibitem[{{Schinnerer} {et~al}\mbox{.}(2007){Schinnerer}, {Smol{\v c}i{\'c}},
  {Carilli}, {Bondi}, {Ciliegi}, {Jahnke}, {Scoville}, {Aussel}, {Bertoldi},
  {Blain}, {Impey}, {Koekemoer}, {Le Fevre}, \& {Urry}}]{schinnerer07}
{Schinnerer} E. {et~al.}, 2007, \apjs, 172, 46

\bibitem[{{Scoville} {et~al}\mbox{.}(2007){Scoville}, {Aussel}, {Brusa},
  {Capak}, {Carollo}, {Elvis}, {Giavalisco}, {Guzzo}, {Hasinger}, {Impey},
  {Kneib}, {LeFevre}, {Lilly}, {Mobasher}, {Renzini}, {Rich}, {Sanders},
  {Schinnerer}, {Schminovich}, {Shopbell}, {Taniguchi}, \&
  {Tyson}}]{scoville07}
{Scoville} N. {et~al.}, 2007, \apjs, 172, 1

\bibitem[{{Sijacki} \& {Springel}(2006)}]{sijacki06}
{Sijacki} D., {Springel} V., 2006, \mnras, 366, 397

\bibitem[{{Sirothia} {et~al}\mbox{.}(2009){Sirothia}, {Dennefeld}, {Saikia},
  {Dole}, {Ricquebourg}, \& {Roland}}]{sirothia09}
{Sirothia} S.~K., {Dennefeld} M., {Saikia} D.~J., {Dole} H., {Ricquebourg} F.,
  {Roland} J., 2009, \mnras, 395, 269

\bibitem[{{Smol{\v c}i{\'c}}(2009)}]{smo09short}
{Smol{\v c}i{\'c}} V., 2009, \apjl, 699, L43

\bibitem[{{Smol{\v c}i{\'c}} {et~al}\mbox{.}(2008){Smol{\v c}i{\'c}},
  {Schinnerer}, {Scodeggio}, {Franzetti}, {Aussel}, {Bondi}, {Brusa},
  {Carilli}, {Capak}, {Charlot}, {Ciliegi}, {Ilbert}, {Ivezi{\'c}}, {Jahnke},
  {McCracken}, {Obri{\'c}}, {Salvato}, {Sanders}, {Scoville}, {Trump},
  {Tremonti}, {Tasca}, {Walcher}, \& {Zamorani}}]{smo08}
{Smol{\v c}i{\'c}} V. {et~al.}, 2008, \apjs, 177, 14

\bibitem[{{Smol{\v c}i{\'c}} {et~al}\mbox{.}(2009){Smol{\v c}i{\'c}},
  {Zamorani}, {Schinnerer}, {Bardelli}, {Bondi}, {B{\^i}rzan}, {Carilli},
  {Ciliegi}, {Elvis}, {Impey}, {Koekemoer}, {Merloni}, {Paglione}, {Salvato},
  {Scodeggio}, {Scoville}, \& {Trump}}]{smo09}
{Smol{\v c}i{\'c}} V. {et~al.}, 2009, \apj, 696, 24

\bibitem[{{Steidel} {et~al}\mbox{.}(1996){Steidel}, {Giavalisco}, {Pettini},
  {Dickinson}, \& {Adelberger}}]{steidel96}
{Steidel} C.~C., {Giavalisco} M., {Pettini} M., {Dickinson} M., {Adelberger}
  K.~L., 1996, \apjl, 462, L17

\bibitem[{{Taniguchi} {et~al}\mbox{.}(2007){Taniguchi}, {Scoville}, {Murayama},
  {Sanders}, {Mobasher}, {Aussel}, {Capak}, {Ajiki}, {Miyazaki}, {Komiyama},
  {Shioya}, {Nagao}, {Sasaki}, {Koda}, {Carilli}, {Giavalisco}, {Guzzo},
  {Hasinger}, {Impey}, {LeFevre}, {Lilly}, {Renzini}, {Rich}, {Schinnerer},
  {Shopbell}, {Kaifu}, {Karoji}, {Arimoto}, {Okamura}, \& {Ohta}}]{taniguchi07}
{Taniguchi} Y. {et~al.}, 2007, \apjs, 172, 9

\bibitem[{{Tasse} {et~al}\mbox{.}(2007){Tasse}, {R{\"o}ttgering}, {Best},
  {Cohen}, {Pierre}, \& {Wilman}}]{tasse07}
{Tasse} C., {R{\"o}ttgering} H.~J.~A., {Best} P.~N., {Cohen} A.~S., {Pierre}
  M., {Wilman} R., 2007, \aap, 471, 1105

\bibitem[{{Trump} {et~al}\mbox{.}(2007){Trump}, {Impey}, {McCarthy}, {Elvis},
  {Huchra}, {Brusa}, {Hasinger}, {Schinnerer}, {Capak}, {Lilly}, \&
  {Scoville}}]{trump07}
{Trump} J.~R. {et~al.}, 2007, \apjs, 172, 383

\bibitem[{{White} {et~al}\mbox{.}(1997){White}, {Becker}, {Helfand}, \&
  {Gregg}}]{white97}
{White} R.~L., {Becker} R.~H., {Helfand} D.~J., {Gregg} M.~D., 1997, \apj, 475,
  479

\bibitem[{{Wilman} {et~al}\mbox{.}(2008){Wilman}, {Miller}, {Jarvis}, {Mauch},
  {Levrier}, {Abdalla}, {Rawlings}, {Kl{\"o}ckner}, {Obreschkow}, {Olteanu}, \&
  {Young}}]{wilman08}
{Wilman} R.~J. {et~al.}, 2008, \mnras, 388, 1335

\end{thebibliography}

\appendix

\bsp

\label{lastpage}

\end{document}